\documentclass[bibyear]{aa}

\usepackage{xcolor}

\usepackage{footnote}

\usepackage{amsmath}
\usepackage{natbib}
\usepackage{float}

\usepackage{booktabs}
\usepackage{graphicx}

\usepackage{txfonts}

\usepackage{hyperref}

\begin{document}

\titlerunning{Near-IR diagnostic diagrams}
   \title{Near-infrared diagnostic diagrams of the gas ionization sources in nearby galaxies: a JWST NIRSpec view}

\author{J. H. Costa-Souza\inst{1,2}
\and Luis Colina\inst{1}
\and Rogemar A. Riffel\inst{2,1}
\and Santiago Arribas\inst{1}
\and Michele Perna\inst{1}
\and Miguel Pereira Santaella\inst{3}
\and Ismael García-Bernete\inst{4}
\and Montserrat Villar Martín\inst{1}
\and Oli L. Dors\inst{5}
}

\institute{Centro de Astrobiología (CAB), CSIC-INTA, Ctra. de Ajalvir km 4, Torrejón de Ardoz, E-28850, Madrid, Spain
\and Departamento de F\'isica, CCNE, Universidade Federal de Santa Maria, Av. Roraima 1000, 97105-900,  Santa Maria, RS, Brazil
\and Instituto de Física Fundamental, CSIC, Calle Serrano 123, E-28006 Madrid, Spain 
\and Centro de Astrobiología (CAB), CSIC-INTA, Camino Bajo del Castillo s/n, E-28692, Villanueva de la Cañada, Madrid, Spain  
\and Universidade do Vale do Paraíba, Av. Shishima Hifumi, 2911, Cep 12244-000, São José dos Campos, SP, Brazil
}
   \date{Received September 15, 1996; accepted March 16, 1997}

  \abstract
   {Tracing the origin of the gas emission lines in galaxies is, in some cases, a difficult task. Buried active galactic nuclei and obscured young stellar clusters pose significant  challenges to traditional optical diagnostic diagrams. Therefore, developing new tools to trace  the excitation sources across the spectrum is a necessary effort for the advancement of the field.}
   {Our goal is to explore the full spectral range of the JWST--NIRSpec data, searching for alternative diagnostic diagrams in the less-explored near-infrared (NIR), and to investigate the nature of the ionizing and heating source. }
   {We analyze the high-resolution spectra of the circum-nuclear regions of nine local ($z < 0.1$) {Ultra/Luminous Infrared Galaxies (U/LIRGS)}, investigating potential emission-line ratios to trace the excitation mechanisms acting on the line-emitting gas. We investigate these objects using the well-established [Fe\,{\sc ii}]\,1.2570$\mu$m/Pa$\beta$ versus H$_2$\,1--0\,S(1)/Br$\gamma$ diagram, and attempt to correlate its classifications with other emission features across the spectrum. We then compare the empirical classifications with \texttt{CLOUDY} photo-ionization models, in order to evaluate how accurately the data can be reproduced. Finally, we compare the 
line width at 80\% of the total flux ($W_{80}$) of selected emission lines with the corresponding gas excitation mechanisms.}
{We propose two line-excitation diagnostic diagrams based on [C\,{\sc i}]\,0.9852$\mu$m/Pa$\gamma$ and H$_2$\,1--0\,O(5)/PAH 3.3\,$\mu$m ratio, which we found to directly correlate  with the hardness of the radiation field, and therefore with the gas excitation mechanisms. Both the empirical [Fe\,{\sc ii}]\,1.2570$\mu$m/Pa$\beta$ versus H$_2$\,1--0\,S(1)/Br$\gamma$ diagram and our proposed [C\,{\sc i}]\,0.9852$\mu$m/Pa$\gamma$ versus H$_2$\,1--0\,S(1)/Br$\gamma$ diagnostic diagram are well reproduced by photo-ionization models. In addition, the line-kinematics analysis shows that the $W_{80}$ values of regions excited by star formation are on average slightly lower 
than those of AGN-excited regions, while shock-excited regions display distinctively higher  $W_{80}$ values compared to the other two groups. Our JWST--NIRSpec results reinforce previous studies showing that the H$_2$ emission in the central regions of AGN hosts is complex and likely produced by multiple excitation mechanisms.}
   {}


   \maketitle
%

\section{Introduction}

The central regions of galaxies can host two distinct phenomena triggered by cold gas accretion: an Active Galactic Nucleus (AGN), driven by the feeding of a supermassive black hole (SMBH), and an starburst event, associated with intense star formation (SF). While both can occur in the same galaxy — and even coexist in its nucleus — they originate from different physical processes. Other mechanisms, such as shocks, can also play an important role in gas excitation and emission, further complicating the interpretation of observational data. Disentangling these phenomena can be challenging, especially when analyzing limited portions of the spectrum. Emission lines, offer rich \textit{in situ} information derived from the line-emitting gas itself, and have long served as a fundamental tool for first classifying supernovae \citep{1968nim..book..483A}; further employed in the study of H \textsc{II} galaxies, \citep{1971ApJ...168..327S}; later implemented to study of AGNs \citep{1981PASP...93....5B,1980A&A....87..152H,1987ApJS...63..295V,2001ApJ...556..121K,2003MNRAS.346.1055K,2005MNRAS.358..363C,2013ApJ...774L..10K,2013ApJ...774..100K}; and U/LIRGs \citep{2010A&A...517A..28M,2006ApJ...637..138M}.  Traditionally, the identification of the gas ionization source relies on emission-line ratios, selected to be close in wavelength,  reducing any biases due to extinction. 

 Analyzing optical emission lines generally sheds light on the uncertainties regarding the ionization source.  However, buried nuclei and dust-enshrouded AGNs \citep{2014ApJ...786..104U,2025A&A...696A.135G} present challenges, as extinction prevents them from being effectively probed in the visible band. The near-infrared (NIR) bands, on the other hand, provide a powerful tool to probe the gas at greater depths, as dust attenuation is much lower than in the optical bands.

Diagrams involving the fluxes of \([{\rm Fe\,\textsc{ii}}]\) and H$_2$ emission lines—specifically \([{\rm Fe\,\textsc{ii}}]\,1.2567\,\mu{\rm m}/\mathrm{Pa}\beta\) and H$_2\,2.122\,\mu{\rm m}/\mathrm{Br}\gamma$—have long been used to probe the line-emitting gas \citep{1998ApJS..114...59L,2002MNRAS.331..154R,2004A&A...425..457R,2012MNRAS.422..252D,2013MNRAS.430.2002R,2015A&A...578A..48C,2021MNRAS.503.5161R,2023A&A...679A..80C}.  The first ratio was found to be particularly sensitive to shocks, correlating also with visible band shock tracers, such as [O {\sc i}] $\lambda$\,6300/H$\alpha$ \citep{1994ApJ...422..521G,1997ApJ...482..747A}. It is believed that shocks and X-ray heating play a major role in the [{\rm Fe\,\textsc{ii}}] excitation \citep{2000ApJ...528..186M,2012ApJ...749..116C}, in which each process yielding a gas within a temperature range, that can be used to distinguish between different ionization sources. Due to the low ionization potential of [Fe {\sc ii}] lines (16.20\,eV), its emission traces partially ionized regions, a region in transition between the fully ionized and neutral gas zones.

The molecular H$_2$ emission, on the other hand, is believed to originate from regions different from those of [ Fe\,\textsc{ii}], since the gas requires cooler temperatures. Observational evidence for this comes from the distinct line broadening observed between the two species \citep{2005MNRAS.364.1041R,2013MNRAS.430.2002R}. Although H$_2$ has a low binding energy of 4.478 eV \citep{1970JMoSp..33..147H}, which makes it very fragile, high-energy photons play an important role in its excitation. For instance, low-energy X-rays are responsible for penetrating dense gas clouds and exciting molecular hydrogen, while Werner and Lyman-band UV pumping redistributes high-energy photons to lower-energy transitions via H$_2$ emission. Shocks can also excite H$_2$, although at much lower velocities ($v$\:<\:25  km\,s$^{-1}$) than those required to excite neutral and ionized gas ($\sim$100\,km\,s$^{-1}$)
\citep{2008ApJS..178...20A,2017ApJ...836...76A,2021MNRAS.504.3265R,2023A&A...675A..86K}.

Polycyclic aromatic hydrocarbon (PAH) emission in the infrared bands can also be used to distinguish between AGN and SF gas emission \citep{2007ApJ...654L..49S,2007ApJ...669..959R,2010ApJ...724.1193O,2014ApJ...790..124S}. PAH features, excited by UV photons mainly from young stars tend to be strong \citep{1999AJ....118.2625R,2012ApJ...746..168D}, on the other hand, in AGN environments its emission is generally dim, or altered due to the harsher radiation fields that can destroy or ionize the molecules \citep{2022A&A...666L...5G,2024A&A...691A.162G}. { Old stellar populations can also excite PAH as demonstrated by \cite{2012MNRAS.426..892G}, and observed by \cite{2024ApJ...962..196O}}. Most PAH features are observed in the mid-infrared, which has limited the number of studies exploring PAH emission as a diagnostic tool for identifying the origin of the gas excitation. In particular, the 3.3\,$\mu$m PAH feature, which originates from neutral PAH molecules, is usually a strong feature in both star-forming (SF) and AGN environments \citep{1989ApJ...345L..59A,2021ApJ...917....3D,2024A&A...691A.162G}, but is usually missed by ground-based NIR observations, as it falls in a spectral region strongly affected by telluric absorptions.

In this paper, we explore the superb quality James Webb Space Telescope (JWST) Near-Infrared Spectrograph (NIRSpec) Integral Field Unit (IFU) observations of nearby galaxies, with the near-infrared spectral range now fully covered (1--5\,$\mu$m), in combination with theoretical predictions, in order to further investigate gas excitation mechanisms and explore different tracers to construct diagnostic diagrams. This paper is organized as follows: Section~2 describes the data and the emission-line profile fitting. In Section~3, we present the photo-ionization model grid used to reproduce the NIR line emission. Section~4 contains the results, along with a brief summary of how the objects have been classified in the literature. In Section~5, we discuss the data and the findings. Finally, Section~6 summarizes the main results and conclusions.

\begin{table*}[h!]
\caption{Sample of galaxies.}
\centering
\begin{tabular}{lcccccccc}
\hline \hline
Object               & Redshift& Distance  & RA       & Dec. & PID &log(L$_{IR}$) & BPT&Scale\\ 
                    &         &(Mpc)  & (deg)    &(deg)&   &(L$_{\odot})$& --&pc/arcsec  \\ \hline \hline
VV114             & 0.020067&84$\,\pm$ 5  & 16.9466 & -17.5070& 1328 &11.71$^{a}$& HII galaxy$^{c}$&409  \\
IRASF14378-3651    & 0.067637& 302$\,\pm$ 21 & 220.2459 & -37.0755& 1204&12.23$^{a}$& Seyfert\,2$^{e}$&1465   \\
IRASF17208-0014         & 0.042810& 189$\,\pm$ 13 & 260.8415 & -0.2835& 1204& 12.46$^{b}$& HII galaxy$^{c}$&916   \\
IRASF23365+3604 & 0.064480&280.37$\pm$ 19  & 354.7554 & 36.3523& 1204&12.20$^{b}$& Composite$^{c}$&1359  \\
NGC3256 North            & 0.009354& 45$\,\pm$ 3  & 156.9635 & -43.9039& 1328&11.64$^{a}$& HII galaxy$^{d}$&221  \\
NGC3256 South           & 0.009354& 45$\,\pm$ 3  &  156.9635 & -43.9054& 1328&--&--&--  \\
Arp220            & 0.018398& 83$\,\pm$ 5  & 233.7384 & 23.5037& 1267&12.28$^{a}$&LINER$^{c}$&403  \\
NGC6240            & 0.024307&107$\,\pm$ 7  & 253.2453 & 2.4009& 1265&11.93$^{a}$& LINER$^{c}$&523  \\
NGC7469            & 0.016268& 66$\,\pm$ 4 & 345.8151 & 8.8739& 1328&11.65$^{a}$ &Syfert\,1$^{c}$&322  \\ \hline
\end{tabular}
\tablefoot{ The first column lists the object names; followed by the redshift (second column); the distance from the NASA/IPAC Extragalactic Database (\nolinkurl{https://ned.ipac.caltech.edu/}; third column); the Right Ascension (RA) and Declination (Dec.) in the fourth and fifth columns, respectively; Proposal ID (PID) in the sixth column; The logarithm of the infrared luminosity between 8--1000\,$\mu$m (\cite{2023ApJS..265...37Y}$_{a}$, \cite{2019ApJS..244...33J}$_{b}$); Seventh column show the type of emission by optical diagnostic diagrams (\cite{2010ApJ...709..884Y}$_{c}$ \cite{2020AJ....159..167L}$_{d}$, \cite{2021A&A...646A.101P}$_{e}$); Column eight displays the projected scale for the observations}.

\label{tab:process}
\end{table*}
\section{Data and Measurements}

We use publicly available data of nearby galaxies obtained with JWST/NIRSpec \citep{2022A&A...661A..80J} using its Integral Field Spectroscopy mode \citep{2022A&A...661A..82B}. To explore potential emission line ratios in the NIR band and investigate the gas excitation mechanisms, we targeted galaxies with full high-resolution NIR coverage ($\sim$1--5\,$\mu$m), albeit the detector gap. Our sample consists of nine galaxies located at distances $\lesssim 300$\,Mpc, listed in Table~\ref{tab:process} along with their respective Proposal IDs and dominant type of emission, based on optical diagnostic diagrams.  Naturally, our sample is highly biased towards  Ultra/Luminous Infrared Galaxies, which may host either a starbursts, AGN or both; as most of the projects were carried out to explore the emission in these targets hidden behind dust. The data were processed through the \textsc{jwst pipeline} version 1.14.0 \citep{2023zndo...6984365B}, using \texttt{jwst-1242.pmap} reference configuration, the spaxels size is 0.1\,$\times$\,0.1 arcsec$^2$, sourced from the Mikulski Archive for Space Telescopes (MAST).

To ensure accurate fits of the emission lines, we integrate the spectra within an aperture of 0.20\,$\arcsec$ radius, slightly larger than the NIRSpec-IFU angular resolution {Full Width at Half Maximum (FWHM)} of $\sim$\,0.15\,$\arcsec$ at 5\,$\mu$m; \citep{2024A&A...690A.171P}, following the procedure described in Appendix~\ref{sec:Honeycomb}.

These measurements are used to investigate the relationships between the different diagnostic diagrams presented in this work. In addition, we use spaxel-by-spaxel measurements to map the gas emission distribution and excitation in individual targets, allowing for comparison with previous results when available.

Before fitting the emission lines, we subtracted the contribution of the underlying stellar population from each spaxel. We use the \textsc{pPXF} code \citep{2017MNRAS.466..798C}, together with the \textsc{E-MILES} stellar population models \citep{2016MNRAS.463.3409V}, to fit the absorption lines and continuum of the spectra. Our fits are restricted to the wavelength range 1--2.6\,$\mu$m, using complementary additive and multiplicative polynomials of degree 3 to correct for the continuum shape. The reason for not fitting the entire spectral window is the high-degree polynomial that would be required to model all molecular features beyond 2.6\,$\mu$m, which could affect the resulting fits at shorter wavelengths. In addition, at wavelengths beyond this range, the contribution of the stellar population to the observed continuum is negligible, as the emission is dominated by dust and molecular features \citep[e.g.][]{2025A&A...696A.135G}. {Thus, at wavelengths longer than 2.6\,$\mu$m, we extrapolate the stellar population fit and model the remaining spectrum using a spline function to account for the observed continuum emission arising from dust grains.}

All the galaxies of our sample, show signs of complex gas motion, likely due to the presence of galactic winds, or interaction with close companions. Thus, to accurately measure the properties of the line-emitting gas in the integrated apertures spectra, we use Levenberg-Marquardt method from \texttt{astropy}, to fit two Gaussian profiles to each emission line using the continuum-subtracted spectra. In addition, we include a first-degree polynomial to account for any residual local continuum resulting from imperfect stellar fitting.  We included an additional broad component to fit the nuclear recombination lines in NGC\,7469 to account for emission from the Broad Line Region (BLR), as it is the only Type 1 nucleus in our sample; this BLR component was only used at the inner apertures, since the BLR is spatially unresolved at the JWST angular resolution.

As we are only interested in line fluxes, widths and its respective uncertainties, we focus on fitting the emission line profile, which according to the region and object, took one or two components.We use the moment-1 map as the initial guess for the Gaussian centroid (calculated over a window of 1200\,km s$^{-1}$ centered at the line wavelength), allowing some tolerance in the fitting parameter space so that the fitter can converge to the optimal solution.  As for the other parameters, we made educated guesses based on the specific characteristics of the galaxy. {To demonstrate the performance of the two-Gaussian component fitting in galaxies with complex gas kinematics, we present representative examples of emission-line fits in Figs.~\ref{fig:F1}, \ref{fig:F2}, and \ref{fig:F3}.} We derive the uncertainties by performing $N=100$ Monte Carlo (MC) simulations. In each simulation, the observed spectrum is perturbed according to the uncertainties provided by the reduction pipeline. The process generates a collection of possible gaussian arrangement, fitted according with the randomic noise inserted in the data. Then we derive the uncertainties of each parameter, through the standard deviation (STD) of each of the possible parameters produced in each MC try. We measure the flux in each MC try by direct integration, estimating the flux uncertainties also using the flux STD of all the tries.
{In addition we measured the flux of the PAH\,3.3\,$\mu$m feature by fitting it with a Drude profile, and estimated its uncertainties through the aforementioned MC procedure.}

Figure~\ref{fig:maps} shows the diagnostic diagrams of log$_{10}$([Fe \textsc{ii}] 1.2570 $\mu$m/Pa$\beta$) versus log$_{10}$(H$_2$ 1--0 S(1)/Br$\gamma$), along with the corresponding 
excitation map, for each galaxy. We estimate the emission line fluxes using spaxel-by-spaxel measurements of the moment-0. We calculate it following a procedure similar to that described in \cite{2021MNRAS.500.4730B}, after continuum subtraction the moment-0 can be {obtained by direct integration over the emission line}. We only use this procedure to measure the fluxes of the  $[{\rm Fe\,\textsc{ii}}]\,1.2567\,\mu{\rm m}$, $\mathrm{Pa}\beta$,  H$_2\,2.122\,\mu{\rm m}$ and $\mathrm{Br}\gamma$ emission lines, used in the standard NIR diagnostic diagram \citep{2002MNRAS.331..154R,2004A&A...425..457R}. {We use a S/N threshold of 18 to detect the lines used at this diagram}.

\section{photo-ionization models}

To aid with the interpretation of the data, we used the \textsc{cloudy} Version \textsc{c23.01} code \citep{2023RMxAA..59..327C,1998PASP..110..761F} to generate grids of photo-ionization models for a very young stellar population (2\,Myr) and for a typical {moderate-luminosity} (log($L_{\rm bol}/L_{\rm Edd}$)= $-$1.15) AGN spectral energy distribution (SED). For the stellar population SED, we used an instantaneous star-formation model generated with the \textsc{Starburst99} code \citep{1999ApJS..123....3L}. \textsc{Starburst99} models both instantaneous and continuous star-formation bursts, allow a choice of initial mass function, evolutionary tracks, and stellar atmosphere models. Given the initial metallicity, the stellar population is evolved over time, and its SED is output for specified time steps.  In our models, star formation was governed by a Kroupa Initial Mass Function (IMF) \citep{2002Sci...295...82K}, with slope $\alpha = -2.3$ for 0.1–0.5\,M$_{\odot}$ and $\alpha = -1.3$ for 0.5–100\,M$_{\odot}$. We adopted Padova evolutionary tracks at solar metallicity \citep{1993A&AS..100..647B} and Lejeune \& Schmutz stellar atmospheres \citep{1992PASP..104.1164S,1997A&AS..125..229L}. For the AGN SED, we used the mean SED model for low $L_{\rm bol}/L_{\rm Edd}$ from \cite{2012MNRAS.425..907J}. The SED includes a robust additive AGN component, \textsc{optxagnf} \citep{2012MNRAS.420.1848D}, fitted over 51 objects, yielding three mean SEDs spanning {moderate} (log($L_{\rm bol}/L_{\rm Edd}$)= $-$1.15) to high (log($L_{\rm bol}/L_{\rm Edd}$)= $-$0.03) $L_{\rm bol}/L_{\rm Edd}$ .

Following \cite{2012MNRAS.422..252D}, our models assume iso-density gas shells surrounding the ionizing source. We use gas densities ($n_{\rm H}$) in the range $10^{3}$–$10^{5}$\,cm$^{-3}$, varying the ionizing source intensity through the ionization parameter $\log(U)$ from $-4$ to $-1$. We also include a micro-turbulence of 1.5\,km\,s$^{-1}$, which contributes to the kinetic pressure in the cloud \citep{2021A&A...652A..66P,2024MNRAS.527.8727E}. As we aim to probe very different species, it is important to choose an appropriate stopping criterion. To produce low ionization species, such [C {\sc i}] and  [Fe {\sc ii}], molecular features like PAH and H$_{2}$, we need to reach very low proton fraction (H$^{+}$/(H$_{2}$\,+\,H)) zones, although not necessarily lower temperatures. AGNs produces a higher volume of ionizing photons, hence even at low proton fractions the gas may remain at higher temperatures, making a temperature stopping criteria not as useful. On other hand, using the proton fraction directly as the stopping criteria assures that the calculations always stops at the same point in the cloud. 

We found that by setting the stopping proton fraction to 0.3\%, H$_{2}$ in the AGN models begins to populate the zones observed in Fig.\ref{fig:hist}. In contrast, the young stellar population have a much softer ionizing SED, so the SF models were left to cool down to 80\,K, allowing the model to reach the bulk of hot and warm H$_2$ emission.

The gas-phase abundance is a key ingredient in modeling photo-ionized gas clouds, as it directly affects the number of emitters of each species and, consequently, the line ratios. Assuming solar abundances is not always appropriate, since various processes can alter the overall abundances in extragalactic objects. For example, bursts of star formation can enrich the gas with certain elements, while star formation quench due to sudden energy or momentum injection can reduce them over long timescales \citep{2025MNRAS.536..430L}. Shocks and outflows can erode or sublimate dust grains, releasing metals that are normally locked in the dust \citep{1995Ap&SS.233..111D}. This is reflected in the literature by the wide range of abundance estimates \cite{1998AJ....115..909S,2007MNRAS.382..251D,2023MNRAS.520.1687A}.

Since the ratios [C {\sc i}]/Pa$\gamma$ and [Fe {\sc ii}]/Pa$\beta$ involve only hydrogen and metal ions, changes in element abundances directly affect them. For carbon, we adopt the average abundance $\log(\mathrm{C/H}) = -3.42$ derived from a sample of Seyfert galaxies \citep{2025MNRAS.540.1608D}, a factor 1.4 of the solar value  by \citep{2010Ap&SS.328..179G}. For iron, due to the lack of AGN host abundance estimates for this complex element, we adopt the solar value from \cite{2010Ap&SS.328..179G}, as we do for the remaining elements. For PAHs, we use the grain abundances of the Orion Nebula and rescale the PAH-to-dust ratio $q_{\rm PAH}$ {to the value observed in Andromeda galaxy, 4.5\% \citep{2014ApJ...780..172D}}.

{The treatment of molecular species in modeling is highly complex and depends on a variety of physical and chemical processes. Within the \texttt{Cloudy} framework, the modeling of PAHs includes photoelectric heating and collisional processes \citep{2001ApJS..134..263W}, as well as stochastic heating \citep{1989ApJ...345..230G}. In our simulations, we adopt the default grain size distribution based on the power-law proposed by \cite{2008ApJ...686.1125A}, using 10 bins.  For the H$_2$ molecule, we adopt the large model, which includes a self-consistent treatment of the molecule with thousands of energy levels, as described in \cite{2005ApJ...624..794S}, instead of the default, much simpler three-level model \citep{1985ApJ...291..722T,1996ApJ...468..269D}.}

\section{Results}

{We focus on the analysis of the spatially resolved diagnostic diagrams, examining the prominent structures exhibited by the galaxies and comparing them with information available in the literature. We also analyze high S/N ratio spectra extracted from circular apertures to investigate the relationships between the NIR diagnostic diagrams and other potential NIR spectral indicators, as well as to compare them with theoretical predictions.}

\begin{figure}
   \includegraphics[width=0.99\linewidth]{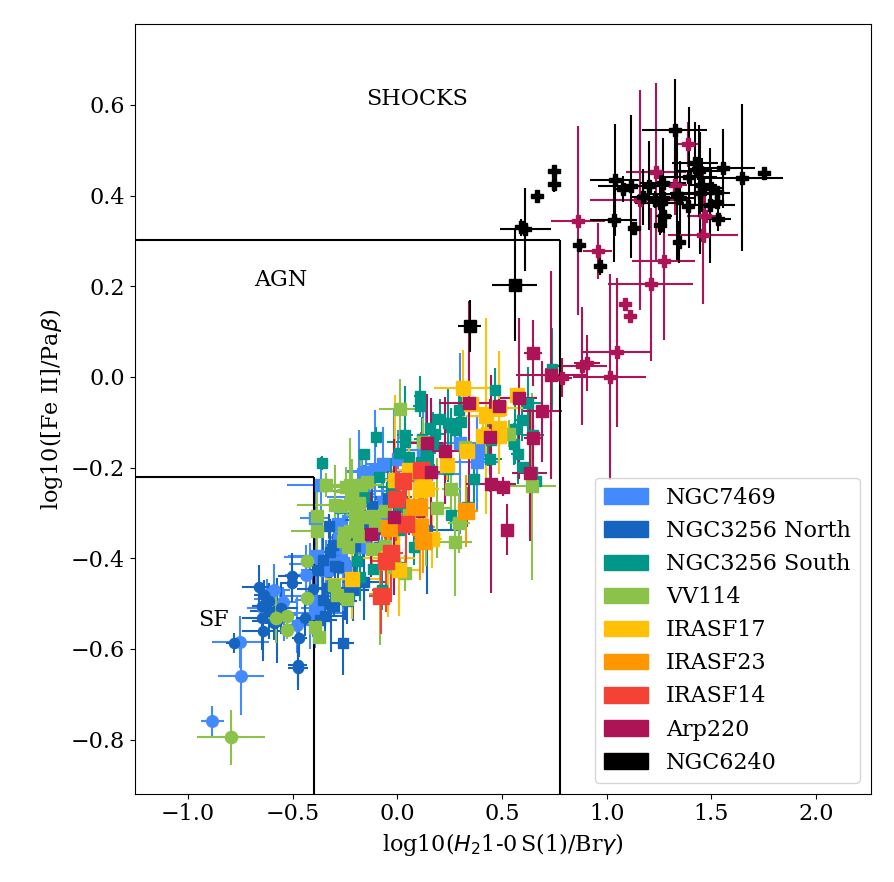}
   \caption{NIR diagnostic diagram based on flux measurements from integrated spectra of all galaxies, using multiple circular apertures. The different galaxies are shown in distinct colors, as indicated in the bottom-right corner of the plot. The black lines indicate the empirical boundaries from \citet{2013MNRAS.430.2002R}.}
   \label{fig:ground}
\end{figure}

{\subsection{Spatially Resolved Diagrams}}

The whole sample is dominated by AGN excitation, although we identify a few specific regions that can be attributed to star formation. This is the case of NGC\,7469, which hosts a well-known circum-nuclear star-forming ring, previously reported by \cite{2023ApJ...942L..36B}, using JWST MIRI and NIRCam images. We also confirm the presence of starbursts in the northern nucleus of NGC\,3256 \citep{2010ApJS..188..447P,2024ApJ...974L..27L}, along with a small star-forming clump in the center of the Southern nucleus, spatially coincident with a dust lane. The NGC\,3256 South nucleus was found to host strong outflows reaching distances of up to 700 pc \citep{2024ApJ...977...36B}, associated to a Compton-thick AGN nucleus \citep{2014A&A...572A..40E}. Our spatially resolved diagnostic maps (Fig.~\ref{fig:maps}) reveal a region consistent with AGN excitation surrounding the northern nucleus. Given the short distance of only $\sim$\,1\,kpc between the two nuclei, we cannot rule out that the gas in the vicinity of NGC\,3256 North is, at least in part, ionized by the southern AGN, either through low-velocity shocks or radiation. In addition, we detect nuclear star formation in VV114, in agreement with previous reports of star-forming sites located in projection near its heavily obscured AGN \citep{2023MNRAS.519.3691D,2024ApJ...966..166B}.

The bulk of the shock-excited regions in the NIR diagnostic diagram are found in only two objects, NGC\,6240 and ARP\,220, both of which present highly disturbed kinematics. In NGC\,6240, AGN emission is observed co-spatially with both confirmed nuclei, while shocks dominate the surroundings of the nuclear regions, in agreement with previous results also based on NIRSpec data \citep{2025A&A...695A.116C}, supporting its late merger stage. We also detect AGN emission in ARP\,220, although there is still an open question in the literature regarding the mechanisms responsible for the line-emitting gas \citep[][and references therein] {2023ApJ...943..142C,2024A&A...690A.171P}. This system is likewise classified as a late merge, given the nuclei proximity (400\,pc), merge induced shocks are very favored \citep{2017ApJ...836...66S}. In addition, most of the gas may be associated with outflows launched by the two nuclei \citep{2020A&A...643A.139P,2025A&A...693A..36U}. Thus, the shocked regions indicated by the diagram are consistent with either, or a combination of the aforementioned events. 
\vspace{.5cm}
\subsection{Line ratios in high-S/N aperture spectra}

Figure~\ref{fig:ground} shows the H$_2$\,1--0\,S(1)/Br$\gamma$ versus [Fe\,\textsc{ii}]\,1.257\,$\mu$m/Pa$\beta$ diagnostic diagram obtained using flux measurements within circular apertures. The S/N threshold per spectrum was set to 50. In addition, we adopt a {reduced $\chi^{2}$ limit of 14 per line to remove spurious measurements; the bulk of the data has $\chi^{2} \leq 1$, although a very small fraction (less than 1\%) of high-quality fits present values above 4, as indicated by visual inspection}. Results for each galaxy are shown in distinct colors, as labeled in the figure. The points in the SF region of the diagram are mostly associated with the galaxies NGC\,7469, NGC\,3256-North, and VV\,114, whereas only Arp\,220 and NGC\,6240 show points in the shock-dominated region. The vast majority of the measurements fall within the AGN region of the diagnostic diagram.

  \begin{figure*}
       \centering
       \includegraphics[width=0.90\linewidth]{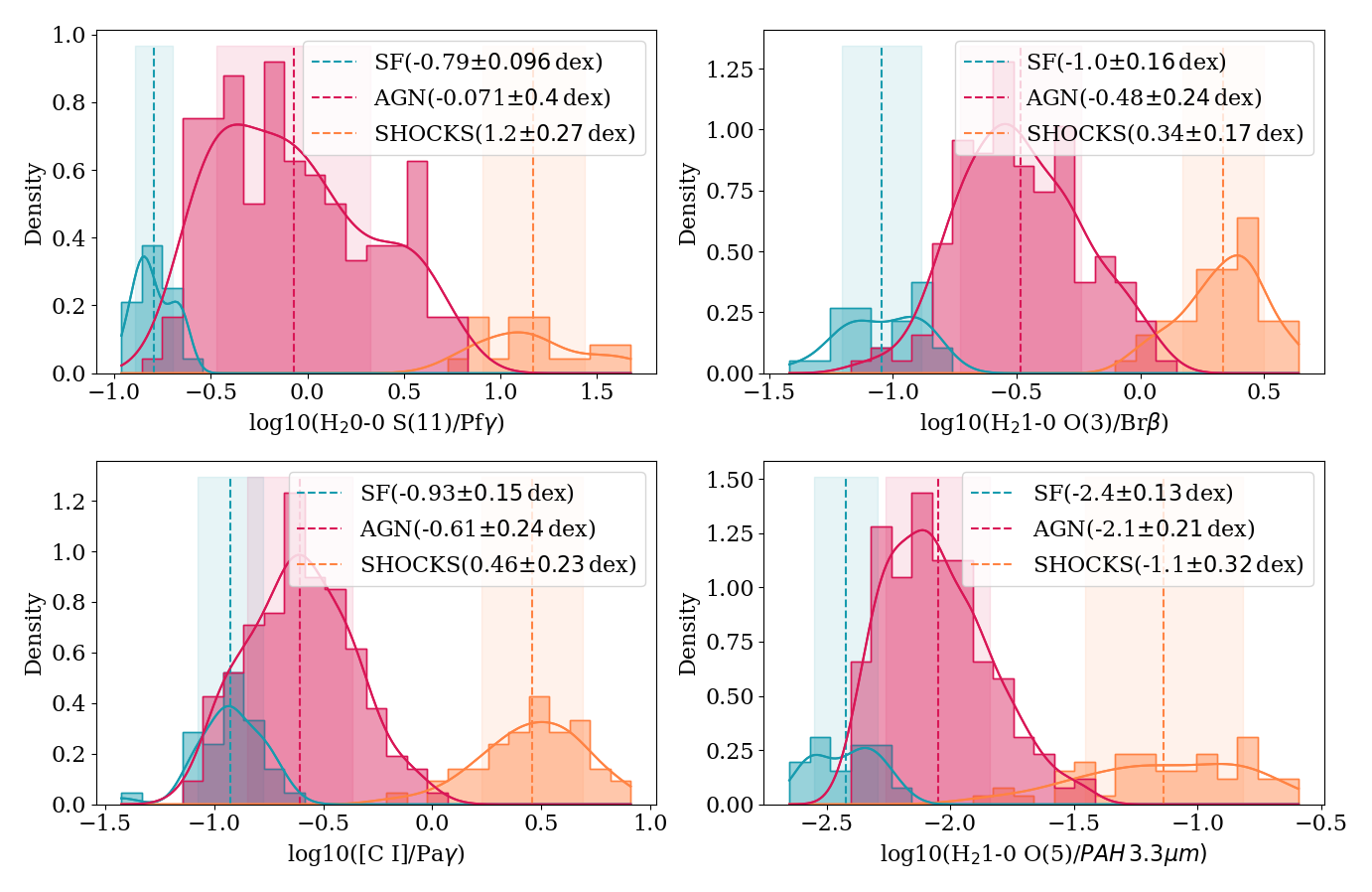}
       \caption{Line-ratio distributions, separated by the different regions of the NIR diagnostic diagram.  {The color-coding is defined according to the regions defined in the H$_2$\,1--0\,S(1)/Br$\gamma$ versus [Fe\,\textsc{ii}]\,1.257\,$\mu$m/Pa$\beta$ diagnostic diagram (Fig.~\ref{fig:ground}): star-forming (in blue), AGN (in red), and shock-dominated (in orange) regions. The solid lines show the interpolation of the distributions.} The dashed lines corresponds to the mean value found for SF (in blue), AGN (in red) and shock excited regions (in orange). The semi-transparent region corresponds to 1$\sigma$ around the average, for each mechanism.}
       \label{fig:hist}
   \end{figure*}

We fitted the strongest emission lines common to all galaxies in the sample (listed in Table~\ref{linetab}) in order to further explore potential diagnostic diagrams aimed at identifying the gas emission mechanisms. Figure~\ref{fig:hist} shows the distribution of $\log$\:H$_2$\,0--0\,S(11)/Pf$\gamma$ (top left panel), $\log$\:H$_2$\,1--0\,O(3)/Br$\beta$ (top right), $\log\:$[C\,{\sc i}]\,0.985$\mu$m/Pa$\gamma$ (bottom left) and $\log$\:H$_2$\,1--0\,O(5)/PAH\:3.3$\mu$m  (bottom right) line ratios, {color-coded according to the regions defined in the H$_2$\,1--0\,S(1)/Br$\gamma$ versus [Fe\,\textsc{ii}]\,1.257\,$\mu$m/Pa$\beta$ diagnostic diagram (Fig.~\ref{fig:ground}): star-forming (in blue), AGN (in red), and shock-dominated (in orange) regions.} The distributions of the AGN- and shock-dominated regions are clearly separated in all four line ratios. In contrast, the SF and AGN regions overlap in all plots; however, as noted above, our sample contains only a few measurements in the SF region.

The ratio of H$_{2}$\,1-0\ O(5) 3.2350$\mu$m to the 3.3$\mu $m PAH behaves similar to the H$_{2}$/Br$\gamma$. It presents values between $-$2.75 to $-$0.59 dex, with the SF regions having the lowest values, reaching up to $-$2.2\,dex; The AGN ionized regions appear between $-$2.41 to $-$1.44, having its lower and higher ends a bit mixed with SF and shock excited regions, respectively; The shocked excited regions present values from $-$1.92 up to $-$0.59 dex. The neutral carbon [C {\sc i}] is one of the weakest features we analyze, though presenting the strongest separation between the AGN and shocked regions, though the worst separation between the SF and AGN ionized. The SF region lies between $-$1.43 to $-$0.67\,dex nearly fully mixed with the AGN ionized points; whilst the AGN points are found between $-$1.16 and $-$0.02\,dex; as for the shocked regions, they lie at the range from $-$0.22 to 0.93\,dex, having the bulk of the points around $\sim$\,0.5\,dex. 
{We performed a two-sample Kolmogorov--Smirnov (K--S) test \citep{1986gft..book.....D}, which shows that the distributions of line ratios from regions excited by SF, AGN, and shocks (Fig.~\ref{fig:hist}) are statistically distinct for the various line combinations (p-value $<0.05$).} 

 \begin{figure*}
    \centering
    \includegraphics[width=0.95\linewidth]{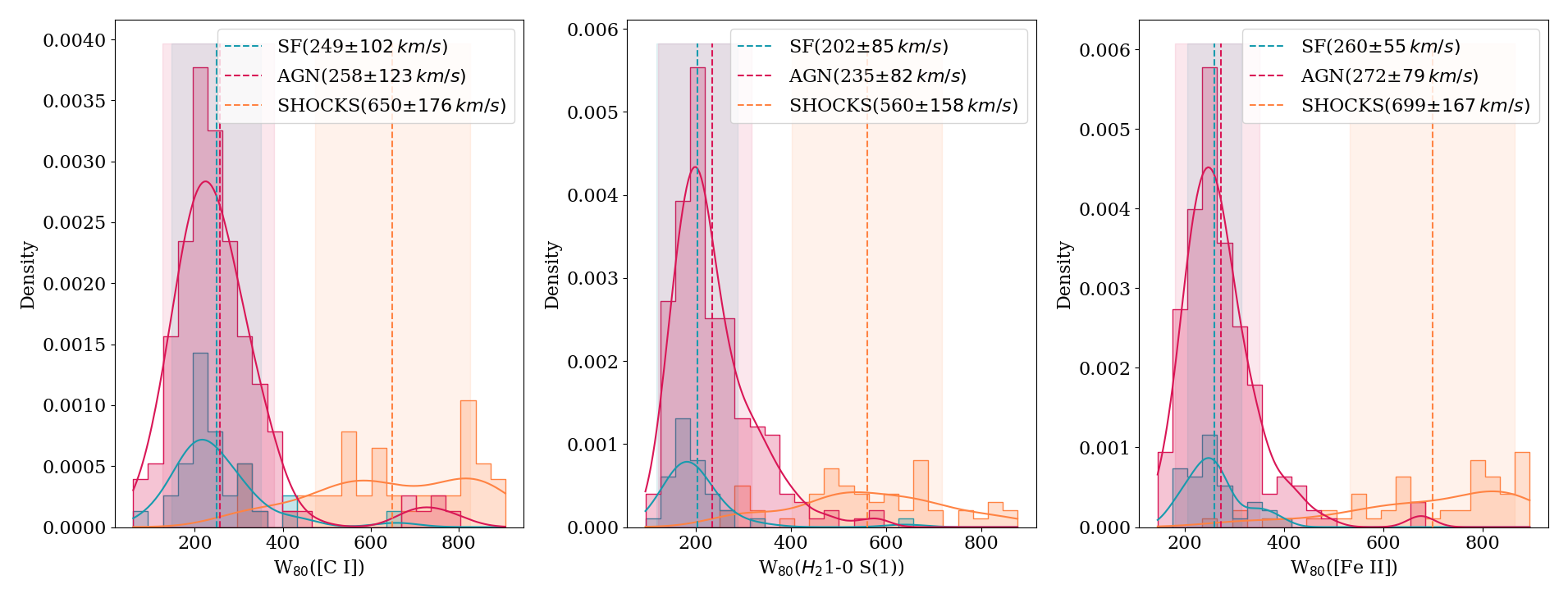}
    \caption{$W_{\rm 80}$ distributions for the whole sample. {The color coding is the same as in Fig.~\ref{fig:hist}.} The three panels show, respectively, the $W_{\rm 80}$ distributions for [C\,{\sc i}], H$_2$\,1-0\,S(1), and [Fe\,{\sc ii}]. The average value of each distribution is indicated by the dashed line and reported numerically in the legend, while the transparent filled region corresponds to the $1\sigma$ dispersion of the distribution. }
    \label{fig:w80}
\end{figure*}
\subsection{Line width}
We also compute the $W_{80}$ parameter for each emission line as a measure of the gas velocity dispersion. This non-parametric indicator is defined as the width of the line that contains 80\% of its total flux and is useful to identify outflows and shock-excited gas \citep{2014MNRAS.442..784Z,2020A&A...642A.147K,2024MNRAS.530.3059G,2021MNRAS.503.5161R,2023MNRAS.521.1832R}.  In Figure~\ref{fig:w80}, we show the [C {\sc i}], H$_{2}$1-0\,S(1) and [Fe {\sc ii}] distributions for our sample, color-coded by the regions of the H$_2$\,1--0\,S(1)/Br$\gamma$ versus [Fe\,\textsc{ii}]\,1.257\,$\mu$m/Pa$\beta$ diagnostic diagram ({Fig. \ref{fig:ground}}).

The highest $W_{80}$ values are found in the shock-excited regions, with mean values  ranging from 560\,$\pm$158\,km\,s$^{-1}$ for the molecular H$_2$ to 700\,$\pm$167\,km\,s$^{-1}$ for [Fe\,\textsc{ii}], for reference the respective median value are 542\,km\,s$^{-1}$ and 750\,km\,s$^{-1}$. The shock-dominated region is clearly separated from the SF and AGN regions, whereas the latter two present similar $W_{80}$ values, with slightly higher mean values seen in the AGN region.  {The two-sample K--S test indicates that, for both ionized gas tracers, the SF and AGN data points are consistent with being drawn from the same distribution, as we cannot reject the null hypothesis (p-value $> 0.05$). In contrast, the test confirms that the shock-excited and photo-ionized gas distributions are distinct (p-value$<0.05$).}

Thus, the $W_{80}$ parameter can be useful for identifying the gas excitation mechanism, particularly when combined with line ratios \citep{2019MNRAS.485L..38D,2019MNRAS.487.4153D,2021MNRAS.503.5161R,2024A&A...682A..71S}.

   \begin{figure*}
       \centering
       \includegraphics[width=0.99\linewidth]{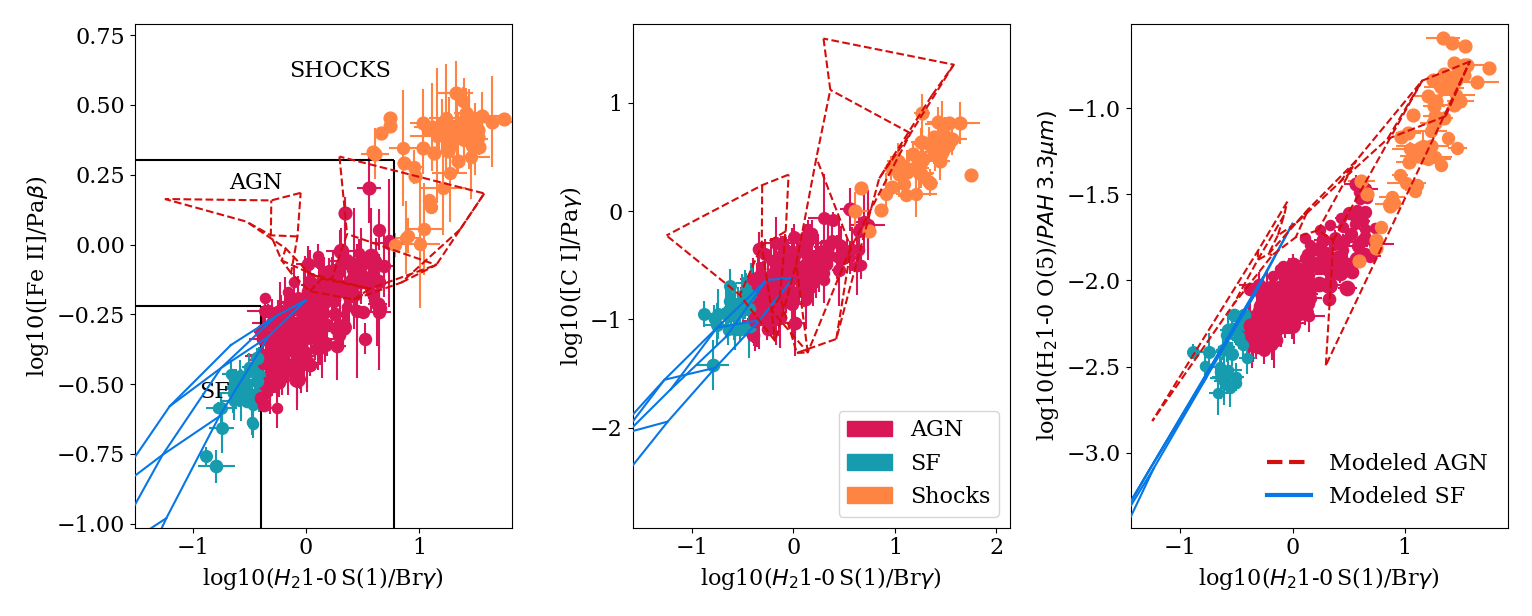}
       \caption{Near-infrared diagnostic diagrams. Left panel show the diagrams [Fe {\sc ii}]/Pa$\beta$ versus H$_2$/Br$\gamma$ from \cite{2013MNRAS.430.2002R}; central panel displays the [C {\sc i}]/Pa$\gamma$ versus H$_2$/Br$\gamma$; at last the right panel show the H$_2$\,1-0\,O(5)/PAH 3.3$\,\mu$m versus H$_2$/Br$\gamma$. The color of the markers indicates the different excitation mechanism based on the left panel: blueish for regions consistent with SF excitation; red, for AGN-excited regions; and orange for shock-excited regions. The AGN photo-ionization models are shown as red dashed lines, while the SF photo-ionization model predictions are presented as blue solid lines. The two first panels show the AGN models to its full extent, while the right panel only display the models with log(U)$\leq$-1.5; in all the panels, only the SF models with log(U)$\leq$ -2.5 are showed. The models densities range from 10$^{3}$ to 10$^{5}$\,cm$^{-3}$. } 
       \label{fig:mod+data}
   \end{figure*}
  
\section{Discussion}

{We explore potential alternative diagnostic diagrams for the line-emitting gas, the physical conditions responsible for its emission, and the relevant literature on their application in Subsection~\ref{sub1}. Subsection~\ref{sub2} is devoted to H$_2$ excitation and the underlying mechanisms, while Subsection~\ref{sub3} discusses the main caveats associated with the proposed diagrams.}

\subsection{Alternative NIR diagnostic diagrams}
\label{sub1}

 We investigated line ratios that could potentially be used to probe the excitation mechanism acting on the line-emitting gas, spread across the NIR and the beginning of the MIR band. In our endeavor to find these line ratios, we may have a look at the spectra produced by the different mechanisms of excitation. In  Fig. \ref{fig:trispec}, we show examples of spectra for the SF-dominated, AGN-dominated and shock dominated regions in NGC\:6240. There is a noticeable difference between the spectrum produced by photo-ionization and shock-excited regions, shock-dominated regions show plenty of strong H$_2$ emission lines, when compared to the other species; the strength of H$_2$ emission is much fainter in the AGN-dominated spectrum, and barely noticeable in locations identified as star formation sites. The 3.3 $\mu$m PAH and hydrogen recombination lines are strong features at the SF-dominated spectrum, however, are not so pronounced at the AGN- and shocks- dominated spectra.

We propose new diagnostic diagrams to trace the excitation mechanisms in {U/LIRGs}. In addition, we test the theoretical limits of the existing [Fe {\sc ii}]/Pa$\beta$ versus H$_2$/Br$\gamma$ diagram as well as those proposed in this work. Figure~\ref{fig:mod+data} shows the sample data, color-coded by excitation mechanism using classifications from \citet{2013MNRAS.430.2002R} (Fig.~\ref{fig:ground}), along with photo-ionization models for AGN- and SF-excited regions (dashed red and solid blue lines, respectively) overlaid on each diagram. For illustration purposes, we limit the range of the photo-ionization models to the observed line ratios. The full extent of the models is shown in Fig.~\ref{fig:placeholder}. 
In the left panel of Fig.~\ref{fig:mod+data} the log$_{10}$([Fe {\sc ii}]\,1.2570$\mu$m/Pa$\beta$) vs. log$_{10}$(H$_2$\,1-0\,S(1)/Br$\gamma$ diagram, the data are well represented by the models in both the SF and AGN ranges. However, some AGN models with higher ionization parameters (log($U$)\,>\,$-$1.5) mimic the behavior of shock-excited gas, slightly overlapping with the shock-dominated region. For the [C\,{\sc i}]/Pa$\gamma$ ratio, there is some overlap between SF- and AGN-excited gas, both in the data and in the models. However, as shown in Fig.~\ref{fig:hist}, this ratio is particularly effective in distinguishing shock excitation from the other excitation mechanisms. {A previous study by \cite{2023A&A...679A..80C} employed the [C\,{\sc i}]\,0.9852\,$\mu$m line in combination with the hydrogen recombination line Pa$\beta$ to distinguish between different ionization sources in the CEERS survey sample \citep{2023A&A...676A..76B}, successfully separating SF- and AGN-dominated regions. When combined with [S\,{\sc iii}] $\lambda\,9530$/Pa$\gamma$, instead of H$_2$\,1--0\,S(1)/Br$\gamma$, the separation in the diagram is significantly enhanced due to the higher ionization potential of the [S\,{\sc iii}] ion (34.86\,eV). In addition, they found some overlap between the photo-ionization and shock models in this diagram.} For reference, we reconstruct the diagram of Fig. \ref{fig:mod+data} in the Appendix (Fig. \ref{fig:reg}), with the limits established in this work and identifying regions of interest from the sample, such as clumps, star forming ring and nuclei.

\begin{table*}[h!]
\centering
\caption{Relationship of the proposed line ratios with the respective excitation mechanism.  }
\begin{tabular}{llll}
\hline
Line Ratio  & SF  & AGN  &Shocks  \\ \hline
log$_{10}$(H$_2$\,0-0\,S(11)\,/\,Pf$\gamma$)     & y\,<\,-0.64 & -0.64\,$\leq$\, y \,$\leq$\,0.8  &  y\,>\,0.8      \\
log$_{10}$(H$_2$\,1-0\,O(3)\,/\,Br$\beta$)       & y\,<\,-0.79 & -0.79\,$\leq$\, y \,$\leq$\,0.06 &  y\,>\,0.06      \\
log$_{10}$([C\,{\sc i}]\,/\,Pa$\gamma$)          & y\,<\,-0.67 & -0.67\,$\leq$\, y\,$\leq$\,0.02  &  y\,>\,0.02      \\
log$_{10}$(H$_2$\,1-0\,O(5)\,/\,PAH\, 3.3$\mu$m) & y\,<\,-2.2  & -2.2\,$\leq$\,y\,$\leq$\,-1.44   &  y\,>\,-1.44      \\ \hline
\end{tabular}
\label{tab:sum}
\tablefoot{The boundaries were taken from the maximum line-ratios produced by the different excitation mechanisms on Fig. \ref{fig:hist}.}
\end{table*}
There are some reports in the literature on the correlation between the H$_2$/PAH ratio and the hardness of the radiation field. For instance, \cite{2020MNRAS.491.1518R,2025ApJ...982...69R} reported a correlation between shock-excited regions and the ratio H$_2$\,0--0\,S(3)/11.3\,$\mu$m\,PAH; \cite{2015MNRAS.451.2640B} found a similar trend; \cite{2010ApJ...724.1193O} observed an increase in the H$_2$/7.7\,$\mu$m PAH ratio in the M\,82 wind region; and radio galaxies were found to show enhanced H$_2$/7.7\,$\mu$m PAH ratios compared to SF galaxies and ULIRGs. \citet{2024A&A...691A.162G} similarly found an anti-correlation between the 6.2\,$\mu$m PAH/H$_2$ 0--0 S(1) ratio and the MIR hardness of radiation field tracers. Further examples can be found in the references therein. Our line ratio (H$_2$\,1-0\,O(5)/PAH 3.3$\mu$m) shows the same trend, it separates SF- of AGN-, as well as shock-excited gas, although with a considerable mixing between the boundaries. The models for the H$_2$\,1--0\,O(5)/PAH 3.3\,$\mu$m ratio {deviate from the data points near the SF–AGN interface, slightly overestimating the ratios; otherwise, they follow the observed data at both ends. The regions we classify as shock-excited can also be reproduced by AGN photo-ionization models with a high ionization parameter}..  This is likely due to the PAH complex nature and abundance diversity of the PAH, for instance, PAH molecules of different sizes have different contribution for each features, e.g. the 3.3\,$\mu$m PAH is heavily due to neutral and small molecules (less than 80 carbon atoms), \cite{1993ApJ...415..397S}.

{Neutral carbon emission is produced at the interface between weakly ionized and molecular gas. Due to its low ionization potential, its emission is a poor tracer of highly excited gas. Instead, the [C\,{\sc i}]/Pa$\gamma$ ratio serves as an indicator of the extent of the ionization zone. PAH emission behaves in a similar way. The 3.3\,$\mu$m feature is attributed to C--H stretching \citep{2003ARA&A..41..241D,2002ApJS..143..455K}, with its emission peaking at the interface between H$_2$ and H$^+$ regions, as noted by \cite{1990ApJ...349..120S}. PAH molecules cannot survive in fully ionized zones, making their emission strongly dependent on shielding by other types of grains. The H$_2$/PAH ratio reflects the degradation of the PAH layer by high-energy photons, while H$_2$ continues to dissipate energy within the gas cloud, exciting the surrounding molecules. Thus, the harder the ionizing radiation field, the higher this ratio is expected to be.}

{In highly shocked environments, carbon emission can be enhanced due to the erosion of carbonaceous grains \cite{2008A&A...492..127S}, although at velocities higher than those at which molecular gas can survive ($v_{s}\geq 50\,\rm km\, s^{-1}$; \cite{1994ApJ...433..797J}), further increasing the [C\,{\sc i}]/Pa$\gamma$ ratio. Small PAH molecules (N$_{C}\,\approx\,50$), to which the 3.3\,$\mu$m feature is primarily attributed, do not survive in environments with $v_{s}\geq 100\,\rm km\, s^{-1}$ \citep{2010A&A...510A..36M}. H$_2$ can be dissociated in less extreme conditions than PAH molecules; however, it reforms more efficiently once the gas has cooled sufficiently, leading to an increase in the H$_2$/PAH ratio \citep[e.g.][]{Guillard09,richings18a,richings18b}. }

Considering all the aspects discussed above, we summarize in Table~\ref{tab:sum} the threshold found to distinguish between SF, AGN, and shock dominated regions, these limits were drawn according with the maximum line ratio found to SF- and AGN- excited gas, respectively corresponding to the boundaries between SF-AGN and AGN-Shocks excitation mechanisms. These relations should, however, be used with caution, particularly due to the low number of SF data points, and in combination with other indicators of gas excitation.

\subsection{The H$_2$ excitation mechanism}
\label{sub2}

\begin{figure*}[]
    \centering
    \includegraphics[width=0.90\linewidth]{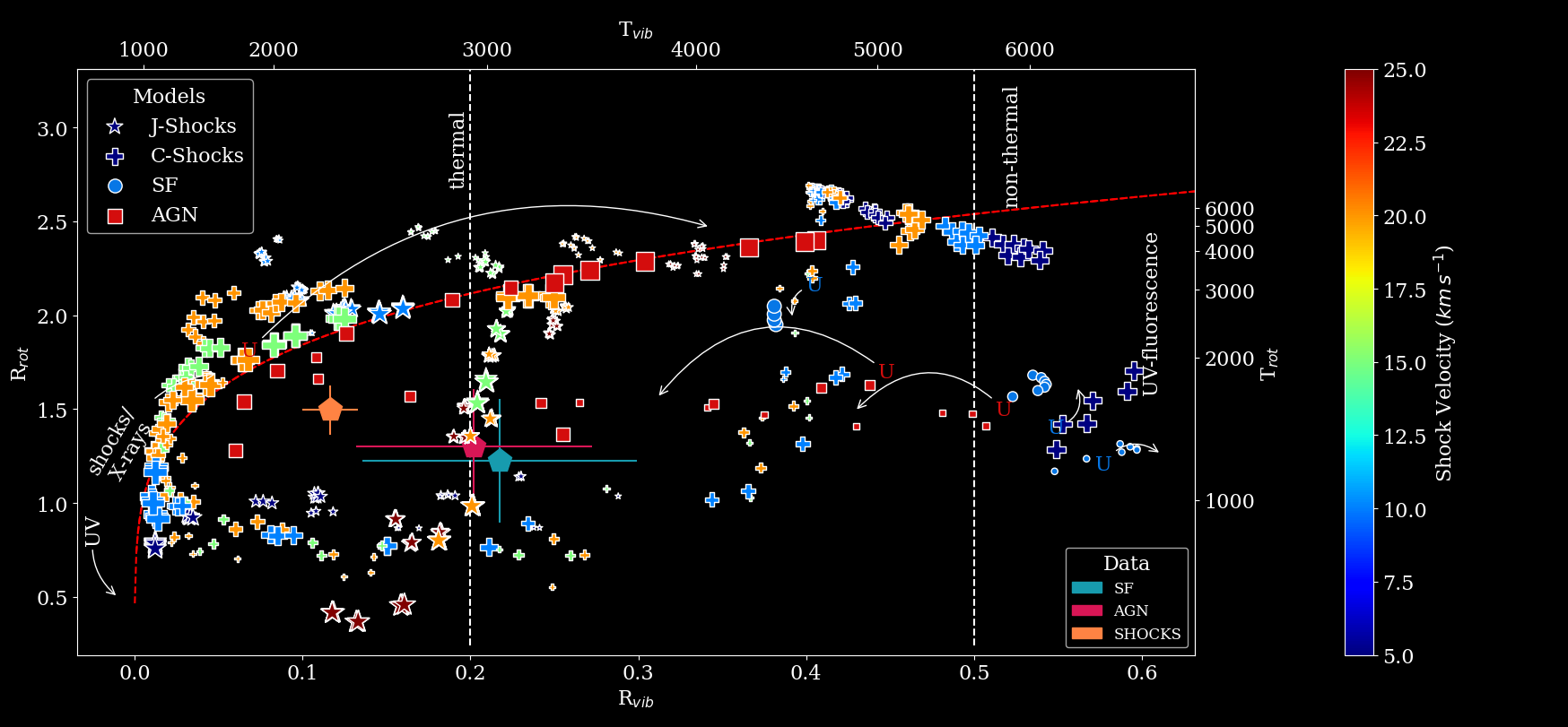}
    \caption{H$_2$ temperature diagram, the bottom x axis shows the ratio H$_2$\,2-1\,S(1)/H$_2$\,1-0\,S(1), and the corresponding vibrational temperature, at the upper axis; The left y axis shows H$_2$\,1-0\,S(2)/H$_2$\,1-0\,S(0), whilst the right axis shows the rotational temperature. {The measured mean line ratios are represented by pentagons of different colors according to the excitation mechanism: starburst (blue), AGN (red), and shocks (orange), with the error bars indicating the standard deviation of the mean values.} The predicted values for both the AGN- and SF- excited regions are respectively displayed as the red-squares and blue-circles. {We also show the J- and C-type shock models from \cite{2023A&A...675A..86K}, represented as stars and crosses, respectively, color-coded according to the shock velocity (blue to red for increasing velocity).} The size of the marker is related to the hydrogen density in all the models, of which the models of SF-excited gas range from 10$^3$--10$^5$\,cm$^{-3}$, AGN-excited gas 10$^3$--10$^6$\,cm$^{-3}$, and the shock-excited gas 10$^3$--10$^7$\,cm$^{-3}$. The ionization parameter at the photo-ionization models, vary from $log(U)$\,= $-$1 to $log(U)$\,= $-$4, white arrows indicates its growth direction. The red dashed line corresponds to the isothermal line.}
    \label{fig:H2EXT}
\end{figure*}

The H$_{2}$ 2--1 S(1)/1--0 S(1) and H$_{2}$ 1--0 S(2)/1--0 S(0) ratios, hereafter referred to as $R_{\rm vib}$ and $R_{\rm rot}$, can provide insights into the dominant excitation mechanism of the H$_2$ gas, through the corresponding vibrational and rotational excitation temperatures (see \cite{1994ApJ...427..777M,2003ApJ...597..907D,2014A&A...572A..40E} and references therein). Fig.~\ref{fig:H2EXT} shows the plot of  $R_{\rm vib}$ vs. $R_{\rm rot}$ for our sample. The observational data are shown as pentagon markers, following the aforementioned color-coding for different excitation mechanisms {(to avoid overcrowding, we display the average position of the data points for regions associated with each excitation mechanism)}; while the red squares show our models for AGN-excitation; blue circles, the young stellar population photo-ionization models; at last the stars {and crosses} represent a subset of the shock models by \cite{2023A&A...675A..86K}, color-coded by the shock velocity. The white dashed lines are a summarization of the data from \cite{1994ApJ...427..777M}, made by \cite{2014A&A...572A..40E}, overall it simplifies the effects in which the gas cloud is under. Thermal radiation is expected to dominate in $R_{\rm vib}$ $\lesssim$ 0.2, as non-thermal effects should present $R_{\rm vib}$ $\gtrsim$ 0.5. As mentioned previously, a few physical mechanisms of the H$_2$ molecule are identified as the main contributors to its excitation:

   The UV pumping mechanism excites H$_2$ molecules by populating vibrationally excited levels through UV radiation at the Werner and Lyman transitions. These higher levels then cascade down via quadrupole emission, with radiative cooling of the upper vibrational levels affecting the 2--1\,S(1)/1--0\,S(1) ratio. Modeling of this process with a non-thermal UV source has shown that, at lower densities ($\sim 10^4$\,cm$^{-3}$), the H$_2$-emitting regions populate low rotational but high vibrational temperatures \citep{1987ApJ...322..412B}. At higher densities, on the other hand, collisional fluorescence dominates, and the emission shifts to lower $R_{\rm rot}$ and $R_{\rm vib}$ values \citep{1989ApJ...338..197S}.
  The radiative cooling of H$_2$ emission depends strongly on the gas kinetic temperature, UV flux, and density. For example, \cite{2003ApJ...597..907D} explored models with varying these parameters. At low temperatures ($T \sim 100$\,K) and low densities ($n \sim 10^3$\,cm$^{-3}$), the results are similar to those of \cite{1987ApJ...322..412B}. At densities of $n \sim 10^{4}$--$10^{6}$\,cm$^{-3}$ and higher temperatures ($T_{\rm max} \approx 1000$\,K), collisional fluorescence dominates, yielding $R_{\rm vib} \sim 0.15$ while $R_{\rm rot} \sim 0.95$. At even higher temperatures ($T \sim 2000$\,K) and densities around $10^4$\,cm$^{-3}$, the vibrational and rotational levels become thermalized.
  At the lowest densities, the star-forming models (presented in Fig.~\ref{fig:H2EXT} by blue circles) populate the region slightly above the models of \cite{1987ApJ...322..412B}. As the density increases, $R_{\rm vib}$ decreases while $R_{\rm rot}$ increases, indicating that vibrational levels are being de-excited through collisions, with the energy redistributed among the rotational levels. {Our data points fall in regions associated with thermal excitation, which appears to be due to the high densities ($\sim 10^4$\,cm$^{-3}$) of the observed regions, combined with a mix of ionization mechanisms such as UV radiation, X-rays, and shocks.}

    The X-ray excitation of H$_2$ shares some similarities with UV excitation. High-energy photons penetrate deep into dense clouds, exciting the gas through photoelectric absorption. The resulting photoelectrons play a key role in dissipating energy via collisions and producing far-UV thermal photons, which induce secondary excitation. H$_2$ emission is largely insensitive to the shape of the X-ray continuum but, similar to UV excitation, correlates with the number of ionizing photons \citep{1996ApJ...466..561M}. At sufficiently high X-ray fluxes, the gas heats faster than it can cool, leading to H$_2$ dissociation and a plateau in the emission. Ultimately, the emission from a cloud illuminated by a X-ray source depends on the interplay between Auger electrons and X-ray-induced photoelectrons, which thermalize with the surrounding gas. Modeling of the H$_2$ emission in NGC\,6240 by \cite{1990ApJ...363..464D} showed that a large fraction of the H$_2$ flux originates from X-ray irradiation of dense clouds ($n \sim 10^5$--$10^6$\,cm$^{-3}$). The models lie in the isothermal region, reflecting the thermalization of the H$_2$ levels at high temperatures.
 In a pure AGN photo-ionization model (represented by red squares at Fig. \ref{fig:H2EXT}), the emission is driven by a combination of X-rays and UV radiation rather than by X-rays alone. The behavior, in this case, is similar to that of the SF excitation models: at lower densities ($n_{\rm H} < 10^4$\,cm$^{-3}$), the AGN-excited regions populate higher $R_{\rm vib} \sim 0.5$. As the density increases, $R_{\rm vib}$ decreases while $R_{\rm rot}$ remains nearly constant. At higher densities and ionization parameters, the AGN-excited regions approach the isothermal line, with the ionization parameter roughly proportional to the temperature. This reflects the heating of the gas by high-energy photons, with energy dissipated through collisional and thermal processes.

    Shocks are identified as one of the main excitation sources of H$_2$ in galaxies with enhanced H$_2$ emission \citep{2020MNRAS.491.1518R,2025ApJ...982...69R,2022A&A...665L..11P}. Shock-excited H$_2$ gas exhibits a distinct emission profile, which can be seen in the column density distribution across rotational levels. This appears as shallower profiles in the H$_2$ excitation diagram, indicated by a lower power-law index $\alpha$, $dN \propto T^{-\alpha}\,dT$, N is the column density of a H$_2$ rotational level \citep{2017ApJ...836...76A,2025ApJ...982...69R}. To illustrate shock excitation in molecular H$_2$, we use a subset of the extensive model grid from \cite{2023A&A...675A..86K}. Their models cover a wide range of parameters; for this work, we focus on J-type and {C-type} shocks with densities $n_{\rm H} \sim 10^3$--$10^7$\,cm$^{-3}$ ( a general range of densities commonly found at ISM measurements) and shock velocities of 5--25\,km\,s$^{-1}$ (shock velocities higher than 25\,km\,s$^{-1}$ are known to dissociate the H$_2$ molecule), considering only models without exposure to an external radiation field.  In the diagram of Fig.~\ref{fig:H2EXT}, {J-shocks models} exhibit a notable behavior. The stars are color-coded by shock velocity. Most low-density regions populate high rotational temperatures ($R_{\rm rot} \approx 2.5$), which is related to much higher rotational temperatures than we see in our data. As density increases, indicated by the larger marker sizes, the gas thermalizes. However, at both extremes of shock velocity, an inversion is observed. For the lowest shock velocities, higher-density regions appear collisionally thermalized, while at lower densities, reduced collision rates allow particles to move more freely and energetically, resulting in stronger collisions. Conversely, at the highest shock velocities, particle velocities are sufficient to thermalize the gas even at lower densities. As the environmental density increases, individual collision energies decrease, which is reflected in a decline of $R_{\rm vib}$. {C-shock models (shown as crosses) are mainly concentrated around the isothermal line, with a smaller number of points at lower rotational temperatures of $\sim$\,1500\,K. Since C-type shock models represent a scenario in which the shock propagates through gas that is already in motion, the H$_2$ molecules possess sufficient thermal energy to efficiently thermalize their energy levels, resulting in the concentration of points around the isothermal line. }

In Fig.~\ref{fig:H2EXT}, the shock-excited regions of NGC\,6240 and ARP\,220 lie on a region closer to the isothermal line, which is dominated by models combining X-rays and UV radiation. Similarly, in Fig.~\ref{fig:mod+data}, photo-ionization models with higher ionization parameters—and thus higher X-ray fluxes—fall within the shock-excited region of the diagram. This is likely because a large fraction of the ionizing continuum produced by high-velocity shocks ($ v\,> 100$\,km\,s$^{-1}$) occurs in the X-ray band \citep{2008ApJS..178...20A}, making it difficult to distinguish from photo-ionization by strong X-ray sources. We will explore potential tracers to break this degeneracy in a future work. From the other sources of excitation, one can infer that the gas lies in an intermediate-density regime, $n_{\mathrm{H}} \sim 10^{4}\,\mathrm{cm}^{-3}$, excited by a combination of collisions and low-velocity shocks ($\sim 25\,\mathrm{km\,s}^{-1}$).

In summary, we conclude that the H$_2$ emission in the SF-dominated region of the NIR diagnostic diagram, besides the SF contribution, includes also a combination of X-ray heating and shocks, as indicated by the models \cite{2003ApJ...597..907D,2023A&A...675A..86K}. The AGN-dominated region is well reproduced by the photo-ionization models, with typical densities in the range $10^{4}$--$10^{5}$\,cm$^{-3}$. Similarly, shock models are able to reproduce the H$_2$ emission in both shock-dominated and AGN regions of the MIR diagnostic diagram, while high-density ($10^{5}$\,cm$^{-3}$) AGN photo-ionization models also fall within the region of the observed line ratios for shock-dominated regions. Thus, the JWST NIRSpec results presented here reinforce previous studies showing that the H$_2$ emission is complex and is likely produced by a combination of excitation mechanisms, such as shocks, young stellar populations and AGNs \citep{2005ApJ...633..105D,2005MNRAS.364.1041R,2013MNRAS.430.2002R,2015A&A...578A..48C,2021MNRAS.503.5161R,2021MNRAS.504.3265R}. In particular, X-rays and shocks seem to play a great role at the excitation of H$_2$ gas in our U/LIRG sample.

\subsection{Diagnostic diagrams: Comparison and
caveats across wavelengths}
\label{sub3}

{The use of the [Fe\,{\sc ii}]\,1.257$\mu$m/Pa$\beta$ vs.\ H$_2$\,1--0\,S(1)/Br$\gamma$ diagnostic diagram to trace excitation mechanisms raises a few issues that may also affect the new diagrams. In particular, we find that regions strongly ionized by X-rays can be easily confused with shock-excited regions, as indicated by the degeneracy between data classified as shock-dominated and our photo-ionization models.  In addition, radiative transfer photo-ionization codes generally provide condensed information corresponding to the spatially integrated emission of the simulated physical scenario. Since we often use spaxels, i.e., small sections of an object, in comparisons with these models, we may be contrasting regions that do not satisfy the same physical conditions assumed in the simulations. Therefore, any comparison between models and spatially resolved (non-integrated) data should be treated with caution. }

{Although the [Fe\,{\sc ii}]Pa$\beta$ vs.\ H$_2$/Br$\gamma$ diagnostic diagram is widely used in the literature for distinguishing between ionization mechanisms \cite{2004A&A...425..457R,2013MNRAS.430.2002R,2015A&A...578A..48C}, it can be strongly affected by projection effects, given the evidence that [Fe\,{\sc ii}] emission does not arise from the same regions as H$_2$ \citep{2005MNRAS.364.1041R,2025ApJ...983...98O}. Thus, aperture and projection effects are expected, even when working at the level of individual spaxels, particularly considering the factor of $\sim$6 in distance between the nearest and the most distant galaxies in our sample. In addition, this diagram relies on weakly ionized gas and ro-vibrational emission lines with excitation temperatures lower than the electron temperatures of 10\,000--20\,000\,K typically found in emission-line galaxies \citep{2017A&A...606A..96P}. This implies that low-excitation species can easily saturate in higher-energy regimes. In contrast, H recombination lines are more persistent and increase with the amount of gas exposed to the ionizing source. Finally, H$_2$ emission can also be reduced due to the dissociation of the molecule by the intense AGN radiation field \citep[e.g.][]{2009MNRAS.394.1148S,2020ApJ...893...80G,2021MNRAS.504.3265R}. As a consequence, excited regions tend to shift toward the lower-left corner of the diagram, where star-forming regions are expected to lie (see Fig.~\ref{fig:ground}). For instance, in the case of X-ray selected AGN, these sources frequently overlap with the locus of star-forming galaxies in the [Fe\,{\sc ii}]/Pa$\beta$ vs.\ H$_2$/Br$\gamma$ diagnostic diagram \citep{2017MNRAS.467..540L,2026ApJS..282...68G}.}

{Another limitation is that there are no shock models that simultaneously cover both molecular and ionized gas, which forces us to compare our empirical data only with photo-ionization models. Molecular and ionized gas require very different energies to be excited: H$_2$ emission is primarily produced by low-velocity shocks \citep[e.g.][]{2023A&A...675A..86K}, whereas [Fe\,{\sc ii}] emission is generally associated with fast shocks \citep[e.g.][]{2000ApJ...528..186M,Pereira-Santaella24b}. As a result, theoretical models do not span the full range of excitation conditions observed, making it challenging to interpret our diagnostic diagrams using shocks alone. We reiterate the need for shock models that incorporate both low- and high-velocity shocks in order to properly investigate the  H$_2$ excitation. This would enable a theoretical framework for diagnostic diagrams that combine ionized and molecular emission lines. At the very least, it is important to quantify the impact of shocks with different velocities on H$_2$ emission, ensuring consistency with existing fast shock models, which currently account only for ionized gas.}

{Finally, our sample is composed of U/LIRGs, whose emission is often attributed to highly obscured active nuclei, nuclear starbursts, or a combination of both \citep{2010MNRAS.405.2505N,2003MNRAS.343..585F}. The complex nature of these ionizing sources may lead to unexpected results if the limits derived in this work are applied to other types of emission-line galaxies. In particular, the frequent mixture of star formation and AGN activity in U/LIRGs may cause our diagram limits to appear tighter than they would be for sources with a single, unambiguous excitation mechanism. The different degrees of dust obscuration affecting each excitation source can also bias diagnostic diagrams toward the least obscured component, as is commonly observed in diagrams based on optical emission lines. Therefore, it is essential to incorporate multi-wavelength information to robustly constrain the origin of the emission and the physical conditions in individual objects. This highlights the importance of spatially resolved data provided by integral field spectroscopy. Therefore, the proposed diagnostic diagrams should be used with caution and, whenever possible, in conjunction with complementary diagnostics.}

\section{Final Remarks}

We used high-quality JWST NIRSpec high-resolution IFU spectra for a sample of nine nearby U/LIRGs ( at projected distances between 45--302\,Mpc), covering the spectral range $\sim$1--5\,$\mu$m, to investigate the gas excitation mechanisms. In addition, we propose new diagnostic diagrams sparsely spread across the NIR band, to distinguish between different excitation mechanisms acting on line-emitting gas: SF, AGN and shocks. The proposed diagrams are very sensitive distinguishing shocks from photo-ionization, they are also weakly affected by extinction, as we use spectrally close emission features. We tested both the new and previously established empirical diagrams against models available in the literature and also compared the gas velocity dispersion, parametrized by the $W_{\rm 80}$ parameter, for regions excited by SF, AGN, and shocks in the sample. Our main results are summarized below:

\begin{itemize}
    \item All objects in the sample show regions with gas excitation consistent with AGN activity, as indicated by the [Fe\,{\sc ii}]/Pa$\beta$ versus H$_2$/Br$\gamma$ diagram. However, a few of these objects also have nuclear regions consistent with SF-excited gas. For instance NGC\,3256-north has no other indicative of the presence of an AGN, though study in other wavelengths revels gas likely shock-excited. Hence, low velocity shocks may enhancement the line ratios [Fe\,{\sc ii}]/Pa$\beta$, H$_2$/Br$\gamma$; { the close proximity to the NGC\,3256 southern nucleus, which hosts an AGN, also strongly influences this result. Since the radiation from the narrow-line region can extend over several kilo-parsecs, it is not surprising to find regions around the southern nucleus being excited by the AGN.}

    \item The comparison of $W_{\rm 80}$ between ionized species ([C\,{\sc i}] and [Fe\,{\sc ii}]) and molecular gas (traced by H$_2$\,1--0\,S(1)) shows that the molecular gas lines are generally narrower than those of the ionized species. The $W_{\rm 80}$ values for SF- and AGN-excited gas are nearly indistinguishable (around 255\,km\,s$^{-1}$ for [C\,{\sc i}], and 266\,km\,s$^{-1}$ for [Fe\,{\sc ii}]). In contrast, shock-excited regions display much broader $W_{\rm 80}$ ($\sim$ 2.5 times the $W_{\rm 80}$ of AGN-excited regions), providing strong evidence that the gas is indeed being excited by shocks. This finding, is also an indication of how sensitive the line ratios [Fe\,{\sc ii}]/Pa$\beta$ and H$_2$/Br$\gamma$ are to shock excitation, separating the photo-ionized/ dynamically cool gas, and the shock/ kinematic disturbed regions in basically two different distributions.

    \item  The photo-ionization models show that, there is a degeneracy at the diagram region populated by shock-excited gas and modeled strong X-ray sources, i.e. high ionization parameter. This is also true for other diagnostic diagrams using [C\,{\sc i}] and the 3.3\,$\mu$m PAH feature. Thus, strong X-rays sources may fall in regions of the diagram, marked as shock excited gas.

    \item The [C\,{\sc i}]/Pa$\gamma$ versus H$_2$/Br$\gamma$ diagram is an effective tracer of shock-excited gas but does not distinguish between AGN- and SF-excited gas. In contrast, the H$_2$/PAH 3.3\,$\mu$m versus H$_2$/Br$\gamma$ diagram can separate all three excitation sources, similarly to the [Fe\,{\sc ii}]-based diagram. As our sample is relatively small and the SF regions are atypical, being also exposed to the high-energy photons produced by the AGN, the boundaries established in this paper need a further confirmation with a larger sample consisting of AGN and starburst dominated galaxies.

    \item In our sample, the main mechanism exciting the H$_2$ clouds is a combination of X-ray heating and shocks. Shock-excited regions lie closer to the X-ray heating models, suggesting that the H$_2$ temperature diagram is more sensitive to the X-ray continuum produced by high-velocity shocks ($v > 100$\,km\,s$^{-1}$) than to the shocks themselves, which also contribute to the gas excitation. Even in SF regions surrounding the AGN, the diffuse X-ray radiation from the AGN appears to play a significant role in the observed H$_2$ emission, injecting enough energy for thermal effects to dominate the gas excitation.

\end{itemize}

\begin{acknowledgements}
We thank the anonymous referee for their critical reading of the paper and valuable suggestions, which helped us improve the manuscript. 
     The data were obtained from the Mikulski Archive for Space Telescopes (MAST) at the Space Telescope Science Institute (STScI), which is operated by the Association of Universities for Research in Astronomy, Inc., under NASA contract NAS 5-03127 for JWST.  JHCS acknowledges the support from Conselho Nacional de Desenvolvimento Cient\'ifico e Tecnol\'ogico (CNPq; Proj. 441722/2023-7) and Coordena\c c\~ao de Aperfei\c coamento de Pessoal de N\'ivel Superior (CAPES;  Finance Code 001). 
     RAR acknowledges the support from CNPq (Proj. 303450/2022-3, 403398/2023-1, \& 441722/2023-7) and CAPES(Proj. 88887.894973/2023-00). LC and SA acknowledge support by grant PIB2021-127718NB-100 from the Spanish Ministry of Science and
  Innovation/State Agency of Research MCIN/AEI/10.13039/501100011033
  and by "ERDF A way of making Europe". MPS acknowledges support under grants RYC2021-033094-I, CNS2023-145506 and PID2023-146667NB-I00 funded by MCIN/AEI/10.13039/501100011033 and the European Union NextGenerationEU/PRTR. MVM research has been funded by grant Nr. PID2021-124665NB-I00  by the Spanish Ministry of Science and Innovation/State Agency of Research MCIN/AEI/10.13039/501100011033 and by "ERDF A way of making Europe". IGB is supported by the Programa Atracci\'on de Talento Investigador ``C\'esar Nombela'' via grant 2023-T1/TEC-29030 funded by the Community of Madrid, and acknowledges support from the research project PID2024-159902NA-I00 funded by the Spanish Ministry of Science and Innovation/State Agency of Research (MCIN/AEI/10.13039/501100011033) and FSE+. 
\end{acknowledgements}

%
%

\bibliography{aa58529-25}

\begin{appendix}

\section{Extraction of the spectra}\label{sec:Honeycomb}
\begin{figure}
\centering
\includegraphics[width=0.99\linewidth]{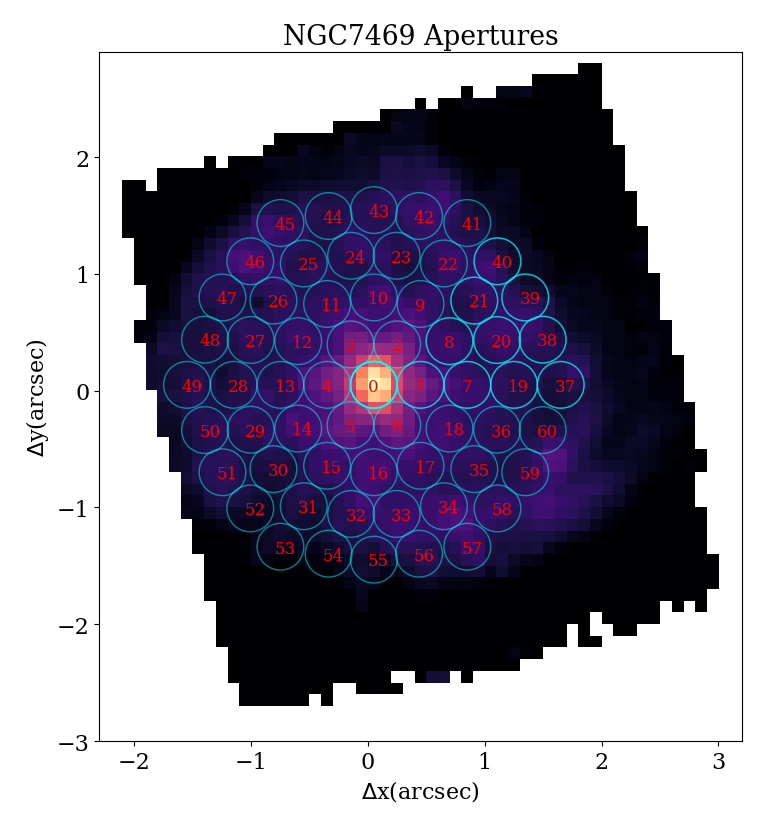}
\caption{Example of the aperture system used to extract the sample's emission line properties.The underlying image is the {2.7\,$\mu$m} continuum image of the galaxy NGC\,7469, and the circles are positioned according to the aperture locations}
\label{fig:Honeycomb}
\end{figure}

Table~\ref{linetab} lists the emission line considered in this work. In Figure~\ref{fig:Honeycomb}, we illustrate the procedure adopted to extract the spectra within circular apertures of 0.20$^{\prime\prime}$ radius. The first aperture was centered on the galaxy nucleus, defined as the position of the continuum peak at {2.7\,$\mu$m} calculated within a spectral window of 120 $\AA$ free of absorption and emission lines. Subsequent apertures were arranged in a honeycomb pattern, as indicated by the numbers in the figure.

\begin{table}
\caption{Line list table}
\centering
\begin{tabular}{ll}
\hline
\hline
Line               & $\lambda_{vac}$($\mu$m) \\ \midrule
\hline
\hline
{[}C I{]}          & 0.9852           \\
He I               & 1.0833\\
Pa$\gamma$      & 1.0941           \\
{[}P II{]}         & 1.1886           \\
{[}Fe II{]}        & 1.2570           \\
Pa$\beta$       & 1.2821           \\
Pa$\alpha$      & 1.8756             \\
H$_{2}$1-0 S(3) & 1.9576           \\
He I             & 2.0518           \\
H$_{2}$1-0 S(2) & 2.0339            \\
H$_{2}$1-0 S(1) & 2.1218            \\
Br$\gamma$      & 2.1661          \\
H$_{2}$1-0 S(0)     & 2.2235          \\
H$_{2}$2-1 S(1) & 2.2477\\ 
Br$\beta$       & 2.6259              \\
H$_{2}$1-0 O(3)  & 2.8025           \\
H$_{2}$1-0 O(5)  & 3.2350           \\
Pf$\gamma$      & 3.7406 \\
H$_{2}$0-0 S(11) & 4.1811           \\\bottomrule
\hline
\end{tabular}
\tablefoot{The vacuum wavelengths for the ionized gas lines were taken from the Atomic Line List\footnote{Atomic Line List v3.00b5, available at \nolinkurl{https://linelist.pa.uky.edu/newpage/}}, while the wavelengths for the H$_2$ are from \cite{2019A&A...630A..58R}.}
\label{linetab}
\end{table}

\section{Examples of line profile fits}\label{app:fits}

\begin{figure*}[b]
    \centering
    \includegraphics[width=0.8\linewidth]{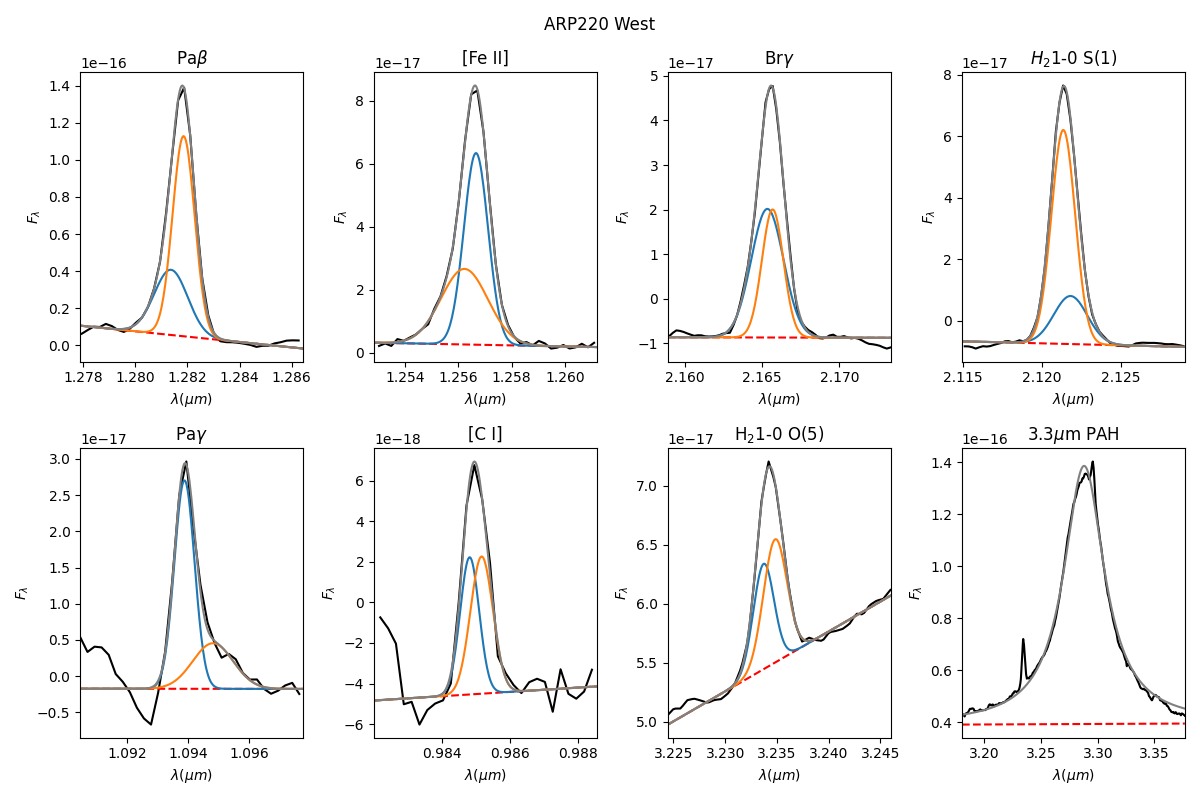}
    \caption{Examples of line fitting for the nuclear aperture of the western nucleus of Arp\,220. We show, respectively, Pa$\beta$, [Fe\,{\sc ii}], Br$\gamma$, H$_2$\,1--0\,S(1), Pa$\gamma$, [C\,{\sc i}], H$_2$\,1--0\,O(5), and the 3.3\,$\mu$m PAH feature. The stellar-subtracted spectrum is shown in black; the two Gaussian components are shown in orange and blue; the underlying first-degree polynomial is shown as a dashed red line; and the sum of all components is shown in gray. For the PAH feature, the single Drude profile is represented by a gray line.}
    \label{fig:F1}
\end{figure*}

\begin{figure*}[b]
    \centering
    \includegraphics[width=0.8\linewidth]{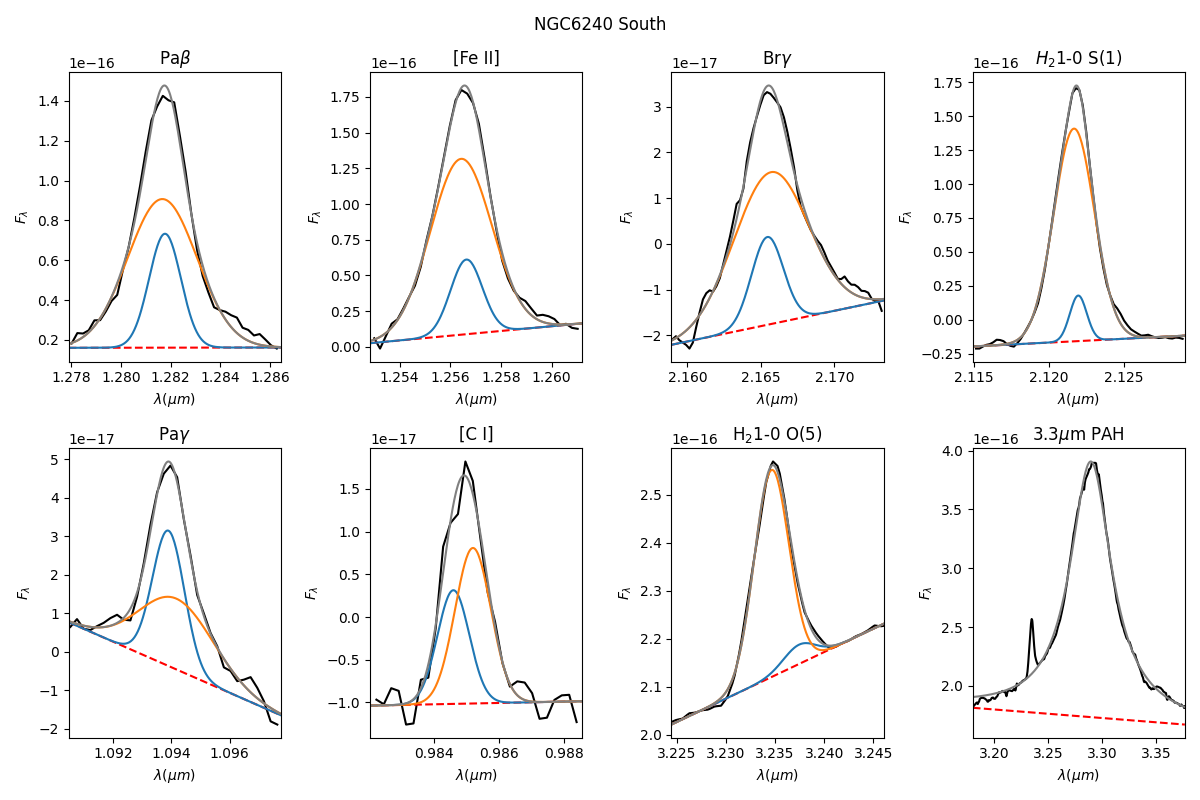}
    \caption{Same as Fig.~\ref{fig:F1}, but for the nuclear aperture of Southern nucleus of NGC6240.}
    \label{fig:F2}
\end{figure*}

\begin{figure*}[]
    \centering
    \includegraphics[width=0.8\linewidth]{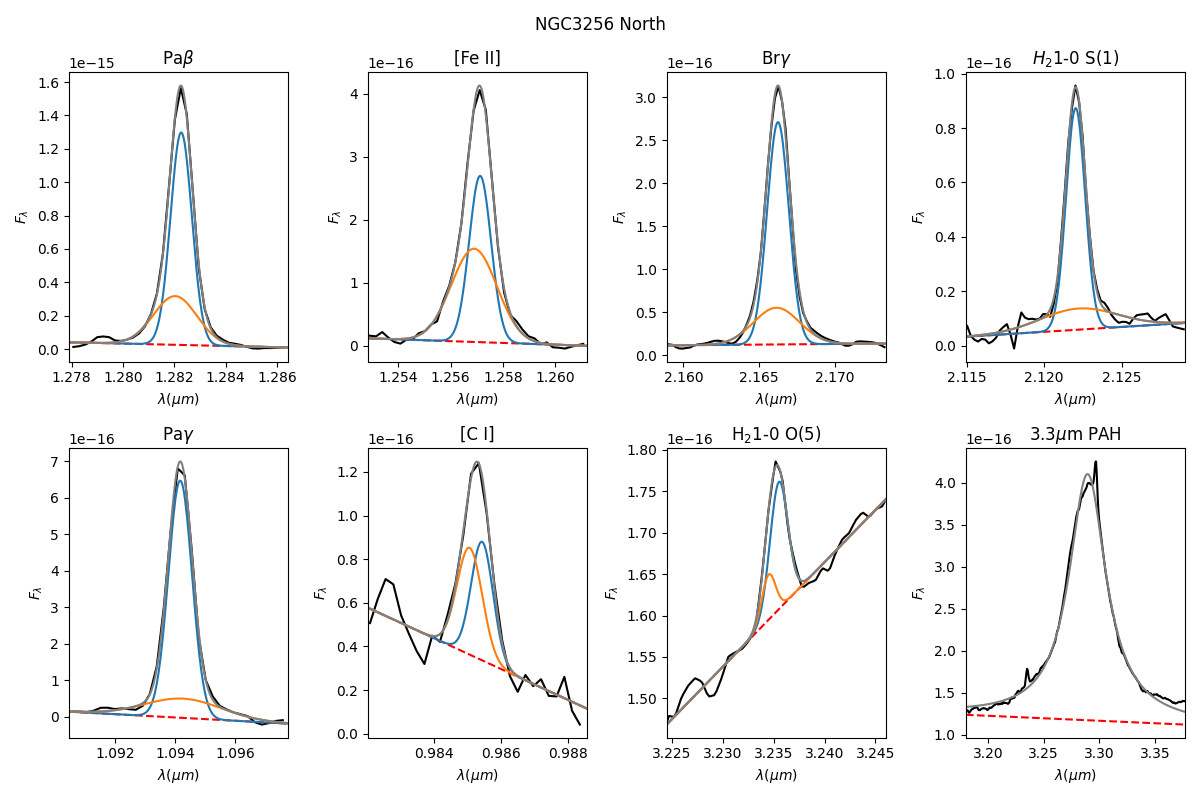}
    \caption{Same as Fig.~\ref{fig:F1}, but for the nuclear aperture of Northern nucleus of NGC3256.}
    \label{fig:F3}
\end{figure*}

{In Figures \ref{fig:F1}, \ref{fig:F2}, and \ref{fig:F3}, we present representative examples of the line fitting for high S/N data, focusing on the most commonly used emission lines. We select two objects with complex kinematics (NGC\,6240 South and Arp\,220) and one with simpler gas kinematics (NGC\,3256 North) to illustrate the use of two Gaussian components to model the line profiles.}

\section{Spatially resolved diagnostic diagrams}
\begin{figure*}
    \centering
    \includegraphics[width=0.73\linewidth]{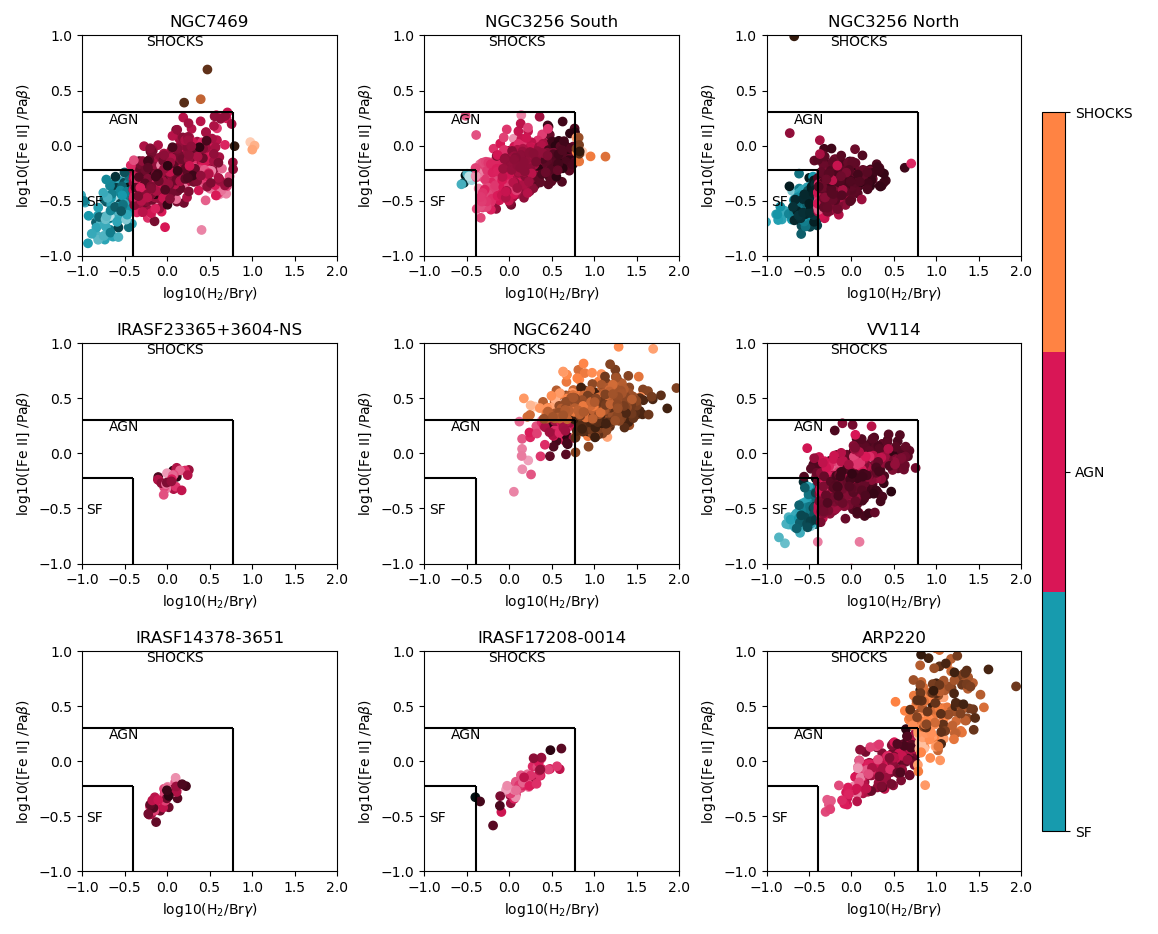}
    \includegraphics[width=0.73\linewidth]{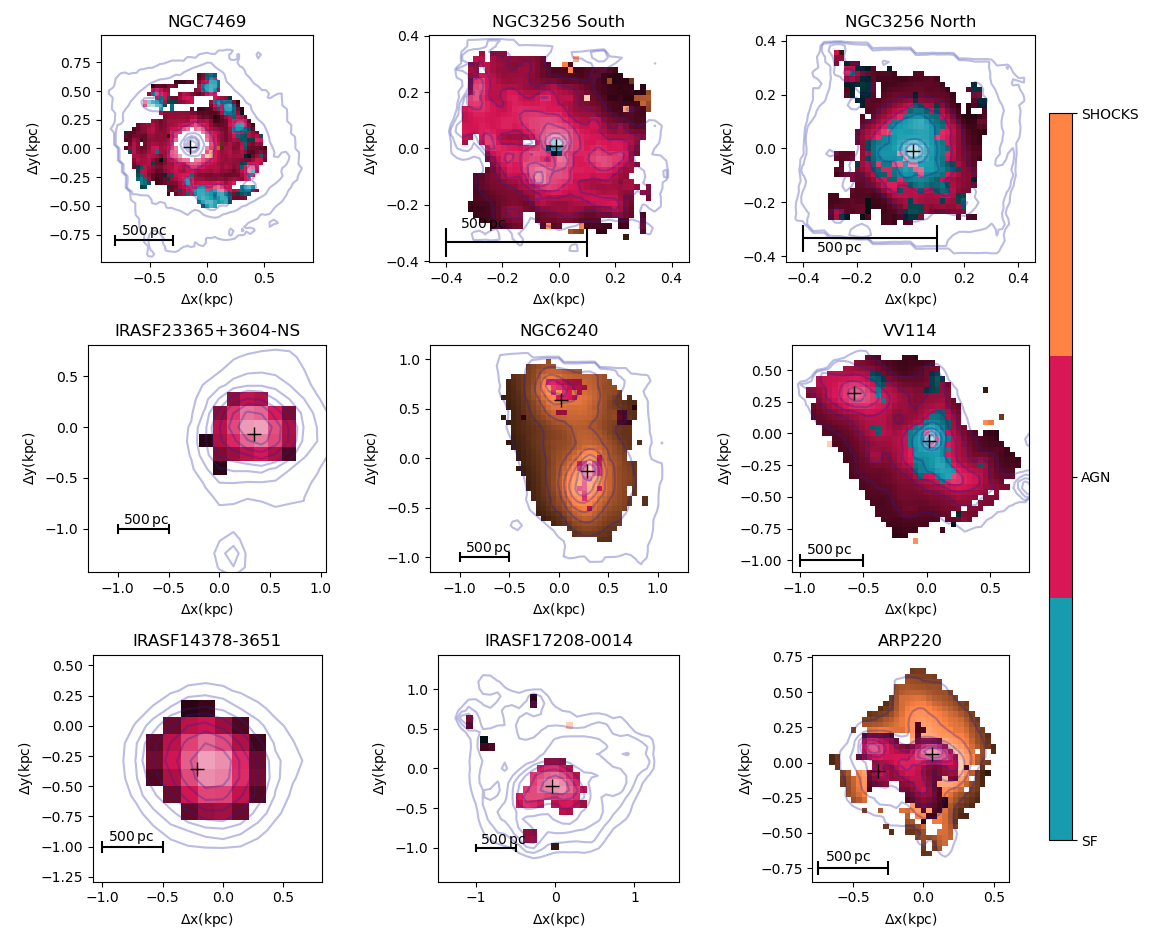}
    \caption{The top block of panels shows the log$_{10}$([Fe {\sc ii}]\,1.2570$\mu$m/Pa$\beta$) versus log$_{10}$(H$_2$\,1-0\,S(1)/Br$\gamma$)  diagnostic diagrams based on spaxel-by-spaxel measurements for each galaxy, while the bottom panels show the corresponding excitation maps. The points are color-coded accordingly the delimiting lines presented in \citet{2013MNRAS.430.2002R}. The black crosses are placed to mark the nucleus in each object, the size of it is the same as the PSF FWHM. We only consider spaxels where the emission lines are detected with $\rm S/N>18$. For reference, we added the contours of the integrated flux at 2.7\,$\mu$m, in semi-transparent blue. We masked the central region of NGC\,7469, as this region is dominated by the emission of the BLR. To analyze the galaxies structures more easily, the brightness of the colors was choosed to be proportional to the integrated flux of Pa$\alpha$ line emission.}
    \label{fig:maps}
\end{figure*}

In Figure~\ref{fig:maps}, we present the H$_2$\,1--0\,S(1)/Br$\gamma$ versus [Fe\,\textsc{ii}]\,1.257\,$\mu$m/Pa$\beta$ diagnostic diagrams for each galaxy, based on the spaxel-by-spaxel flux measurements, along with the corresponding excitation maps. We adopt the separation criteria from \citet{2013MNRAS.430.2002R}, where star-forming regions have H$_2$\,1--0\,S(1)/Br$\gamma$ $<$ 0.4 and [Fe\,\textsc{ii}]\,1.257\,$\mu$m/Pa$\beta$ $<$ 0.6; AGN-excited gas has $0.4 \leq$ H$_2$\,1--0\,S(1)/Br$\gamma$ $\leq 6$ and $0.6 \leq$ [Fe\,\textsc{ii}]\,1.257\,$\mu$m/Pa$\beta$ $\leq 2$; and shock-excited gas has H$_2$\,1--0\,S(1)/Br$\gamma$ $> 6$ and [Fe\,\textsc{ii}]\,1.257\,$\mu$m/Pa$\beta$ $> 2$. The analysis of the gas excitation mechanisms of {9} objects using NIRSpec data has already been in part presented in the literature: 

\begin{itemize}
    \item \textit{NGC7469}, optically classified as hosting a type 1 nucleus (see Table \ref{tab:process}), exhibits most of its points in the AGN region of the diagnostic diagram, with points falling in the SF region only at locations along the circum-nuclear star-forming ring, consistent with the results reported by \citet{2024ApJ...965..103B};
    
    \item The nuclear region of \textit{VV114} is dominated by star formation, while the surroundings are consistent with AGN excitation. This corroborates with previous report of star formation sites at close projected to the highly obscured AGN \citep{2024ApJ...966..166B} and also with the results found by \citet{2023MNRAS.519.3691D}, revealing that VV114 southwestern nucleus emission is likely produced by a stellar cluster, while there is evidence of a highly obscured AGN in the northeastern nucleus; 
    
    \item The NIR diagnostic diagram for \textit{IRASF14378-3651} only presents points in the AGN region, but restricted to the unresolved nucleus. This galaxy was previouly classified as a Low Ionization Nuclear Emission Region (LINER), based on mid-infrared observations \citep{2022MNRAS.517.4162P};
    
    \item \textit{IRASF23365+3604} presents only points in the AGN region of the diagram, but the excitation map is restricted to the very central region (inner $\sim$200 \,pc). Previous optical classification studies placed this galaxy as a composite source (Table \ref{tab:process});
    
    \item \textit{IRASF17208-0014} being one of the most distant galaxies in the sample, has a quite faint emission limited to the central few spaxels, which is compatible with AGN excitation,consistent with previous works indicating a buried Compton thick nucleus, in X-ray \citep{2011A&A...529A.106I,2009ApJ...704.1570G}, MIR \citep{2022A&A...663A..46G} and sub-mm observations \citep{2021A&A...651A..42P};
    
    \item \textit{NGC\,3256}, formed by the collision of two separate galaxies, hosts an starburst in its northern nucleus and a type 2 Seyfert in its southern nucleus \citep{2014ApJ...797...90S,2014A&A...572A..40E,2021ApJS..257...61Y}. The excitation map for the southern counterpart of NGC\,3256 shows mostly AGN-dominated regions, while the northern (main) galaxy exhibits SF line ratios within the inner $\sim$ 200\,pc radius, with higher line ratios, particularly H$_2$\,1--0\,S(1)/Br$\gamma$, observed at larger distances from the nucleus. Previous analysis of the NIRSpec data for this system by \citet{2014A&A...572A..40E,2024ApJ...977...36B} revealed the presence of warm molecular gas outflows from the southern nucleus, while no outflows are detected in the nothern nucleus. The enhancement of the H$_2$/Br$\gamma$ ratio is likely produced by shocks associated with the outflows, as discussed by these authors.

    \item \textit{NGC6240} Most of the line ratios for NGC\,6240 fall in the shock-dominated region of the NIR diagnostic diagram, except at the locations of the two nuclei, where the line ratios are consistent with an AGN origin. MIR line ratio diagnostic diagrams also reveal the presence of shock-excited gas emission in NGC\:6240\,\citep{2025A&A...693A.321H}. Our diagnostic diagram and excitation map are similar to those presented by \citet{Ceci25}, based on the NIRSpec data of this system, leading to the conclusion that circum-nuclear gas excitation is dominated by shocks, while the emission from the two nuclei is associated with the AGN ionization cones;

    \item \textit{Arp\,220} shows line ratios in both the AGN and shock-dominated regions of the NIR diagnostic diagram. The AGN-like values are observed mostly along the east-west direction, covering both the brightest nucleus, and the stellar cluster \citep{2024A&A...690A.171P} above the faintest one, while shock-dominated regions are found at locations farther away. Previous works using NIRSpec data of Arp\,220 do not presented the NIR diagnostic diagram, but discussed the origin of the gas emission, concluding that Arp\,220 host massive SF sites, as well as outflows. MIR analysis points at least one of Arp\,220 nuclei as a compact obscured nucleus \citep{2022A&A...663A..46G}. Although it still lacks direct evidence of hosting an AGN \citep{2024ApJ...977...55G}, in any of its nuclei, and it could be explained by massive SF \citep{2025A&A...693A..36U,2024A&A...690A.171P}.

\end{itemize}

\section{Examples of spectra}

\begin{figure*}
    \centering
    \includegraphics[width=0.97\linewidth]{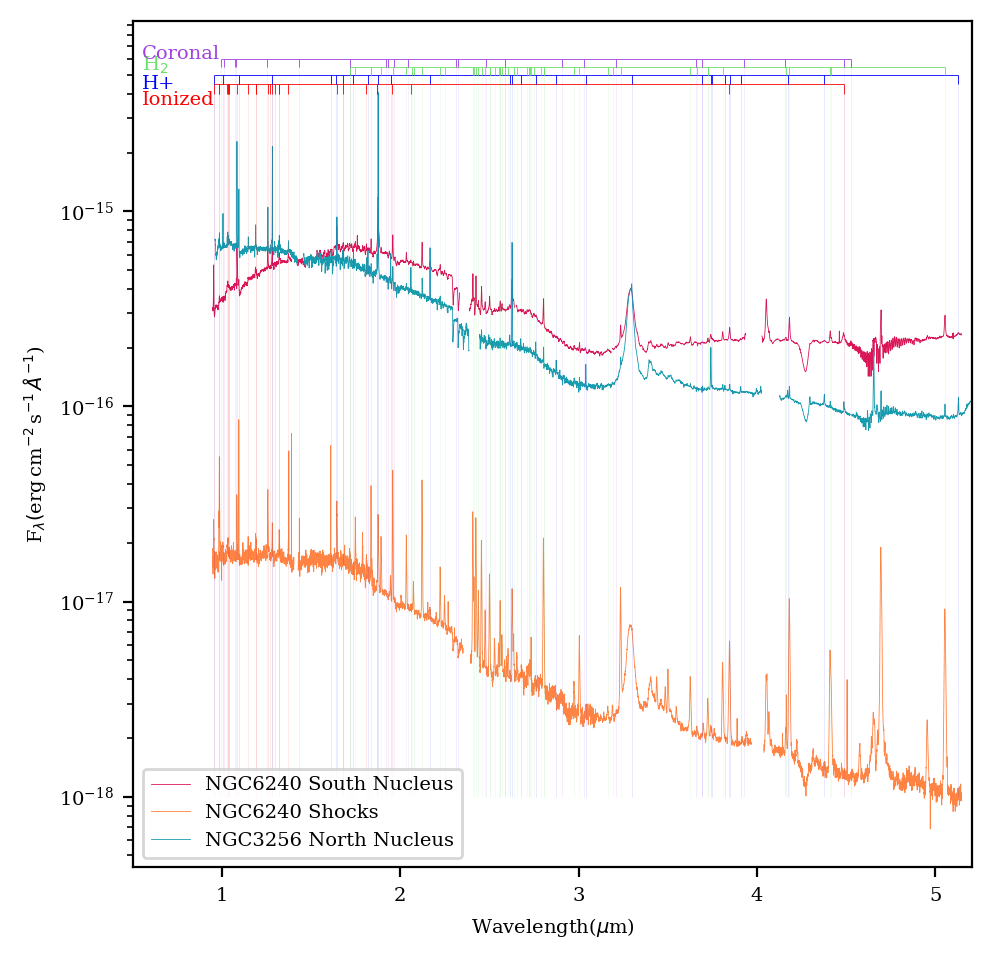}
    \caption{Spectra example for: SF-excited gas at the main nucleus of NGC\,3256 (light blue); Shock-excited gas, spectra of a region between NGC\,6240 nuclei (orange); AGN-excited gas, NGC\,6240 South nucleus (red). }
    \label{fig:trispec}
\end{figure*}

Figure~\ref{fig:trispec} displays examples of spectra from shock-, SF-, and AGN-excited regions, shown in orange, blue, and red, respectively. For reference, the positions of different types of line emission (H$_2$, hydrogen recombination, ionized gas, and coronal lines) are indicated at the top of the plot.

\section{photo-ionization models}

\begin{figure*}
    \centering
    \includegraphics[width=0.99\linewidth]{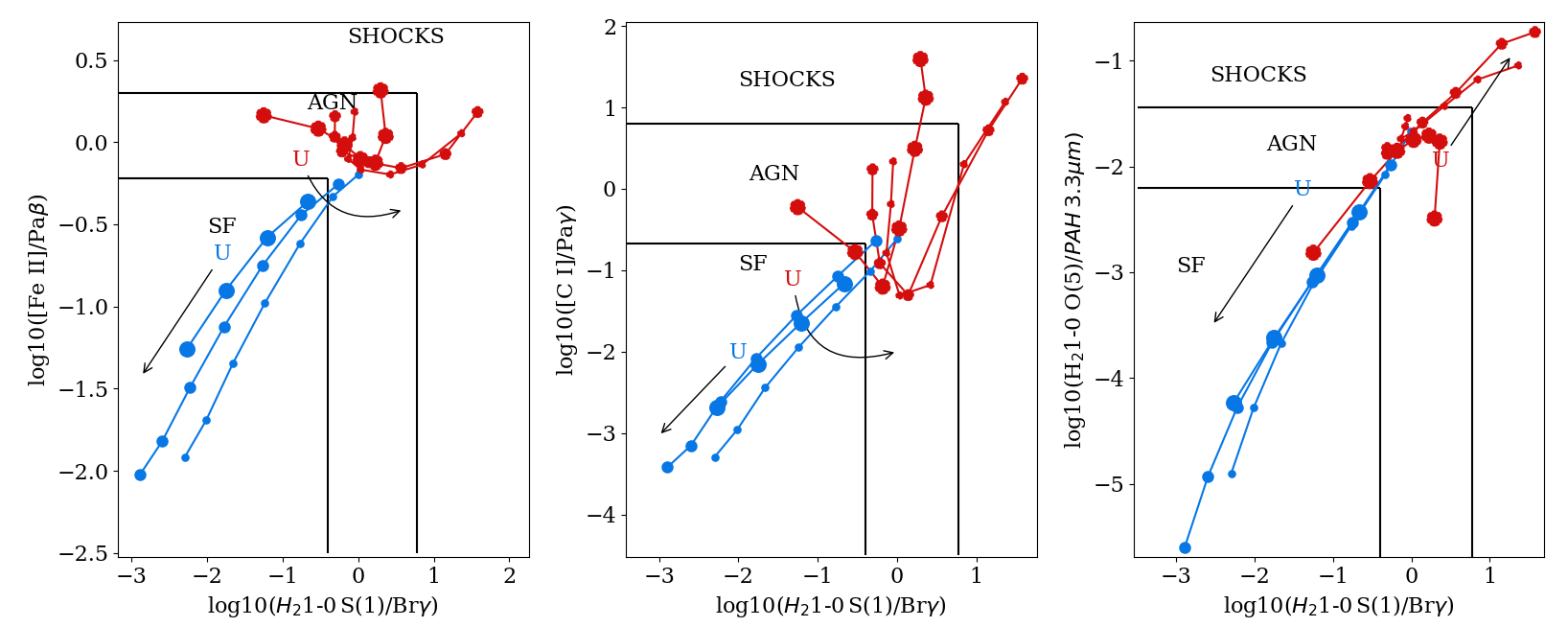}
    \caption{\textsc{Cloudy} models for photo-ionization by AGN (red) and young stellar populations (blue). Arrows indicate the direction of increasing ionization parameter ($U$) from -4 to -1, on a 0.5\,dex step. Marker size represents the different hydrogen densities ($n_{\rm H}$) adopted,  10$^3$,10$^4$ and 10$^5$cm$^{-3}$. The black lines outline the regions proposed to separate the different excitation mechanisms, the limits can also be found in Table \ref{tab:sum}}. 
    \label{fig:placeholder}
\end{figure*}

Figure~\ref{fig:placeholder} shows the full range of line ratio values obtained from the photo-ionization models, with AGN models shown in red and SF models in blue.

\section{Additional diagnostic diagrams}

\begin{figure*}
    \centering
    \includegraphics[width=0.99\linewidth]{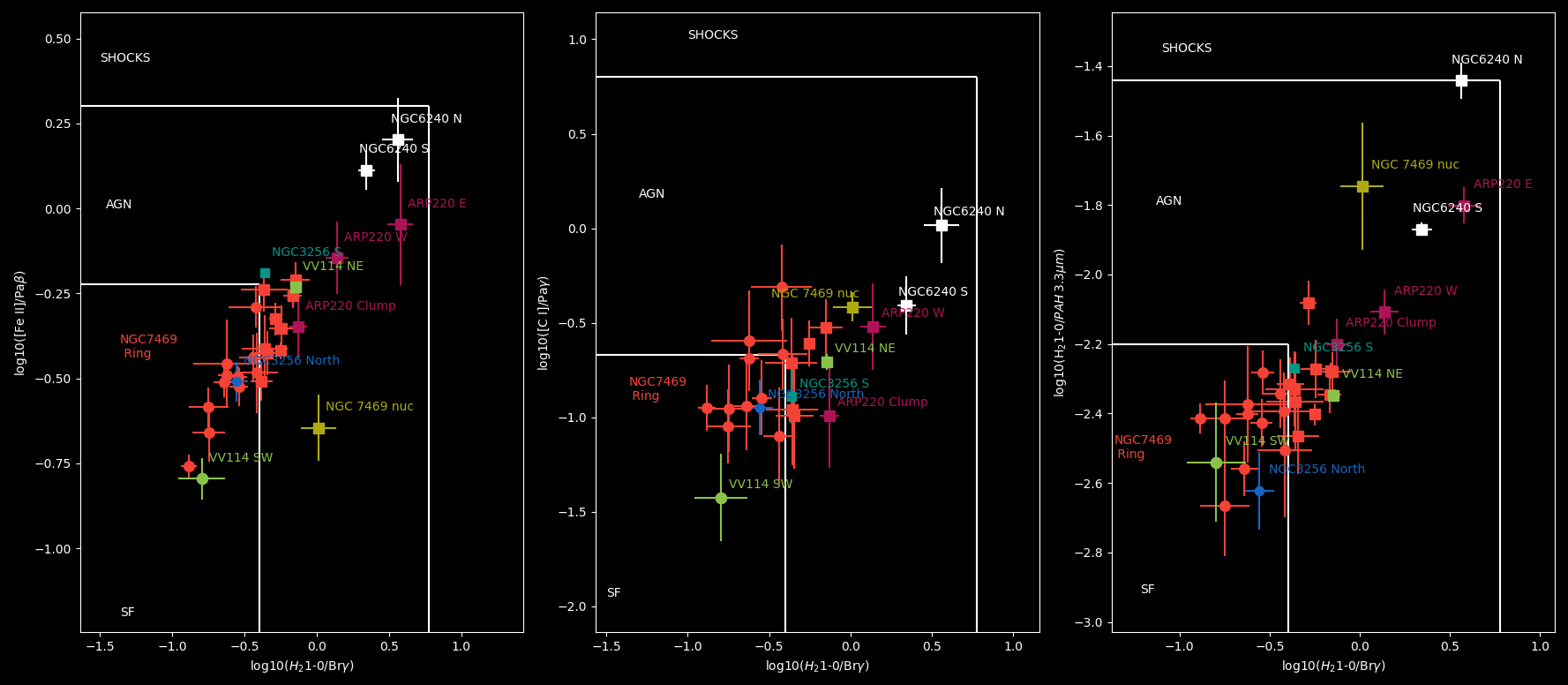}
    \caption{Near-infrared diagnostic diagrams. Left panel show the diagrams [Fe {\sc ii}]/Pa$\beta$ versus H$_2$/Br$\gamma$ from \cite{2013MNRAS.430.2002R}; central panel displays the [C {\sc i}]/Pa$\gamma$ versus H$_2$/Br$\gamma$; at last the right panel show the H$_2$\,1-0\,O(5)/PAH 3.3$\,\mu$m versus H$_2$/Br$\gamma$. The color of the markers indicates the different region of interest on the galaxies, e.g. clumps, multiple nuclei, and regions of a stellar ring. We add the limits measured at this work for reference (white lines). The markers indicate the mechanism exciting the region based at the [Fe {\sc ii}]/Pa$\beta$ versus H$_2$/Br$\gamma$ diagram, circles for young stars, and squares for AGN excitation.}
    \label{fig:reg}
\end{figure*}

Figure~\ref{fig:reg} shows the diagnostic diagrams proposed in this work. The markers represent regions of interest within the galaxies, including multiple nuclei, clumps, and rings. The white lines indicate the proposed limits, while different colors are used to label the regions. Circles denote star-forming regions, and squares indicate AGN-excited regions, based on the $\log_{10}$([Fe\,{\sc ii}]\,1.257\,$\mu$m/Pa$\beta$) versus $\log_{10}$(H$_2$\,1--0\,S(1)/Br$\gamma$) diagram (left panel). To avoid overcrowding, only regions from objects exhibiting competing excitation mechanisms are displayed in the figure.

\end{appendix}

\end{document}